\documentclass[fleqn,usenatbib,useAMS]{mnras}

\usepackage{graphicx}	
\usepackage{amsmath}	
\usepackage{amssymb}	
\usepackage{multicol}        
\usepackage{bm}		
\usepackage{pdflscape}	
\usepackage{booktabs}  
\usepackage{multirow}
\usepackage{subcaption}
\usepackage[T1]{fontenc}
\usepackage{ae,aecompl}

\usepackage{newtxtext,newtxmath}

\title[Galaxy Zoo DESI]{Galaxy Zoo DESI: Detailed Morphology Measurements for 8.7M Galaxies in the DESI Legacy Imaging Surveys}

\author[M. Walmsley]{Mike Walmsley$^{1}$,\thanks{Contact e-mail: \href{mailto:michael.walmsley@manchester.ac.uk}{michael.walmsley@manchester.ac.uk}}
Tobias G\'eron$^{2}$,
Sandor Kruk$^{3}$,
Anna M.~M.~Scaife$^{1,4}$,
\newauthor
Chris Lintott$^{2}$,
Karen L. Masters$^{5}$,
James M. Dawson$^{6}$,
Hugh Dickinson$^{7}$,
\newauthor
Lucy Fortson$^{8,9}$,
Izzy L. Garland$^{10}$,
Kameswara Mantha$^{8,9}$,
David O'Ryan$^{10}$,
\newauthor
Jürgen Popp$^{7}$,
Brooke Simmons$^{10}$,
Elisabeth M. Baeten$^{11}$,
Christine Macmillan$^{11}$ 
\\
$^{1}$Jodrell Bank Centre for Astrophysics, Department of Physics \& Astronomy, University of Manchester, Oxford Road, Manchester M13 9PL, UK\\
$^{2}$Oxford Astrophysics, Department of Physics, University of Oxford, Denys Wilkinson Building, Keble Road, Oxford, OX1 3RH, UK\\
$^{3}$European Space Agency (ESA), European Space Astronomy Centre (ESAC), Camino Bajo del Castillo s/n 28692 Villanueva de la Ca\~{n}ada, Madrid, Spain\\
$^{4}$The Alan Turing Institute, 96 Euston Road, London NW1 2DB, UK\\
$^{5}$Departments of Physics and Astronomy, Haverford College, 370 Lancaster Avenue, Haverford, Pennsylvania 19041, USA\\
$^{6}$South African Radio Astronomy Observatory (SARAO), Black River Park North, 2 Fir St, Cape Town, South Africa, 7925\\
$^{7}$School of Physical Sciences, The Open University, Milton Keynes, MK7 6AA, UK\\
$^{8}$School of Physics and Astronomy, University of Minnesota, Minneapolis, Minnesota, 55455, USA\\
$^{9}$Minnesota Institute for Astrophysics,  University of Minnesota, Minneapolis, Minnesota, 55455, USA\\
$^{10}$Department of Physics, Lancaster University, Lancaster LA1 4YB, UK\\
$^{11}$Citizen Scientist, Zooniverse c/o University of Oxford, Keble Road, Oxford OX1 3RH, UK
}

\date{Last updated XXX; in original form XXX}

\pubyear{2023}

\begin{document}
\label{firstpage}
\pagerange{\pageref{firstpage}--\pageref{lastpage}}
\maketitle

\begin{abstract}
We present detailed morphology measurements for 8.67 million galaxies in the DESI Legacy Imaging Surveys (DECaLS, MzLS, and BASS, plus DES). These are automated measurements made by deep learning models trained on Galaxy Zoo volunteer votes. Our models typically predict the fraction of volunteers selecting each answer to within 5-10\% for every answer to every GZ question. 
The models are trained on newly-collected votes for DESI-LS DR8 images as well as historical votes from GZ DECaLS. We also release the newly-collected votes.  
Extending our morphology measurements outside of the previously-released DECaLS/SDSS intersection increases our sky coverage by a factor of 4 (5,000 to 19,000 deg$^2$) and allows for full overlap with complementary surveys including ALFALFA and MaNGA. 
\end{abstract}

\begin{keywords}
catalogues, software: data analysis, methods: statistical, galaxies: bar, galaxies: interaction, galaxies: general
\end{keywords}



\section{Introduction}

Galaxy images reveal diverse structures such as spiral arms, bars, bulges, and tidal features \citep{Buta2013}. The field of galaxy morphology seeks to understand the origins of these structures. Relatedly, these structures are thought to both influence and trace key physical processes in galaxy evolution and so by measuring their presence one can infer the history of those physical processes \citep{Casteels2013, Geron2023} .

Measuring the morphology of large samples of galaxies is crucial because many highly correlated variables influence both morphology and the processes they trace. Unpicking these correlations requires large samples where one can hold these variables fixed and still retain enough galaxies to draw statistically robust conclusions \citep{Masters2019a}.
One may also hope to find rare populations of galaxies with properties that challenge our assumptions about galaxy formation (e.g. \citealt{Smethurst2021,Keel2022}).

The scale of our morphology measurements is limited not by our supply of telescope images but by our interpretation of those images.
Modern astronomical observatories capture detailed images of millions of galaxies - a sample impossible for astronomers to even begin to review by eye.
To meet this challenge, astronomers have developed methods to measure the detailed morphology of galaxies through parametric and non-parametric fitting (e.g. \citealt{Abraham1996,Simard2002,Conselice2003,Lotz2004}), 
citizen science (including Galaxy Zoo e.g. \citealt{Lintott2008, Willett2013}),
or machine learning (e.g. \citealt{Huertas-Company2008,Banerji2010,Ferrari2015}).

Combining deep learning with citizen science can achieve morphology measurements with the classification detail of humans and the scale of automated systems (e.g. \citealt{Dieleman2015, Sanchez2018}, or see \citealt{HuertasCompany2023Dawes} for a review). Galaxy Zoo DECaLS (\citealt{Walmsley2022decals}, hereafter W+22) was the first to present a large-scale catalogue of morphology measurements for every Galaxy Zoo question by training deep learning algorithms on citizen scientist responses. This catalogue covered the 314,000 galaxies imaged by the Dark Energy Camera Legacy Survey (DECaLS) DR5 and within the SDSS DR8 footprint. 

\begin{figure*}
    \centering
    \includegraphics[width=\textwidth]{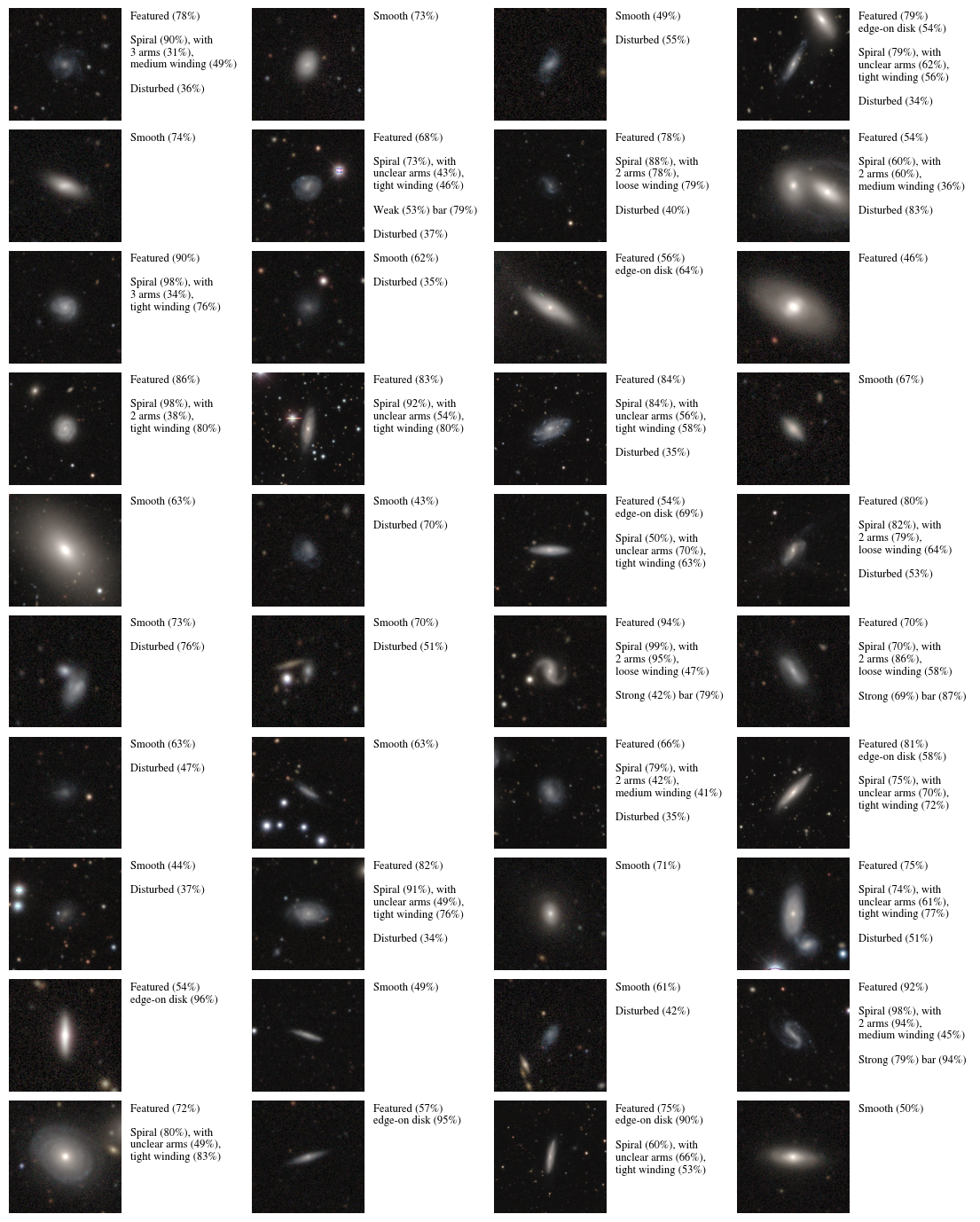}
    \caption{DESI-LS galaxies with their GZ DESI automated morphology measurements. Percentages reflect the percentage of volunteers predicted to select that answer. We show only galaxies where at least 15\% of volunteers are expected to vote `Featured' and with a redshift below 0.1, to better illustrate the detail of our morphology measurements; galaxies are otherwise randomly selected.}
    \label{fig:mosaic}
\end{figure*}

DECaLS is part of the DESI Legacy Imaging Surveys (DESI-LS), a set of three sister surveys designed to produce images with similar characteristics. Here, in GZ DESI, we exploit this imaging similarity to extend and apply the deep learning methods developed for GZ DECaLS to all three DESI-LS surveys. We release new automated predictions for 8.7M bright ($r < 19$) galaxies in DESI-LS DR8. Fig. \ref{fig:mosaic} shows random example galaxies and their automated morphology measurements.

\begin{figure}
    \centering
    \includegraphics[trim={0 0.8cm 0 3cm},clip]{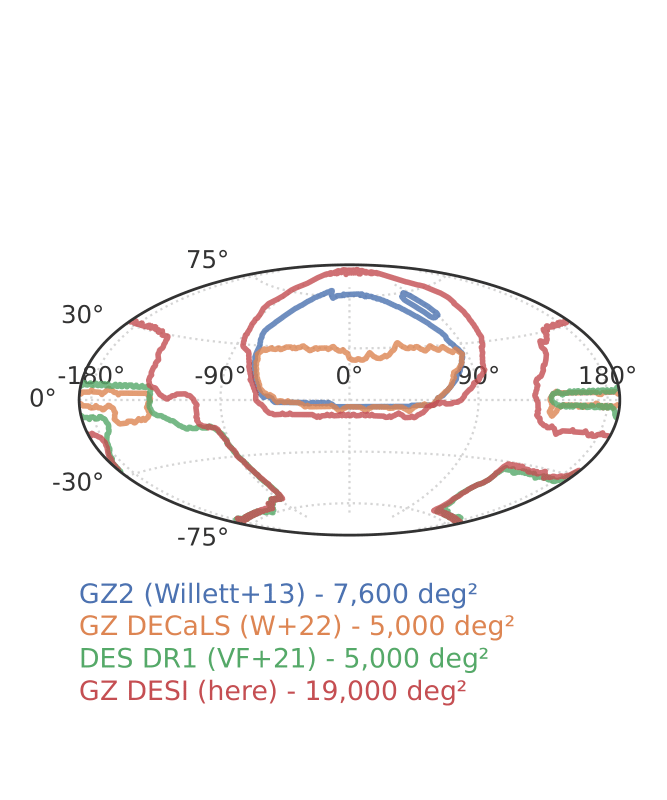}
    \caption{Sky coverage of GZ DESI (i.e. (DECaLS/BASS/MzLS/DES), GZ DECaLS (i.e. the DECaLS and SDSS intersection), and DES in DESI-LS DR8}
    \label{fig:sky_coverage}
\end{figure}

The key benefit of our new morphology catalogue is scale. Including MzLS and BASS, along with additional images from DECaLS not classified in GZ DECaLS, increases our sky coverage from 5,000 deg$^2$ to 19,000 deg$^2$, with a proportional increase in galaxies of all types.
This increase in sample size is crucial for investigating specific morphologies (e.g. weak bars) or controlling for astrophysical variables (e.g. mass, star formation, environment, etc), particularly when constructing volume-limited subsets of well-resolved galaxies. Fig. \ref{fig:sky_coverage} compares our coverage with various existing morphology catalogues and surveys.

Classifying morphology at this scale is made possible by combining citizen science with deep learning.
Our general approach is to train models primarily on the volunteer labels collected during GZ DECaLS and then predict what volunteers would have said for images from DECaLS' sister surveys.
In practice, despite the similarity of the images, making accurate predictions is non-trivial. 
One key complication is label drift \citep{Storkey2013}.
Due to changes in the Galaxy Zoo decision tree, website content, and other factors, volunteer votes collected in our most recent campaign (2020-2022) may not be equivalent to volunteer votes collected at the start of the first Galaxy Zoo DECaLS campaign (2015).
We would like to predict what volunteers might say now, with the current decision tree, and not what they might have said during earlier campaigns.
To benefit from the seven years of Galaxy Zoo labels collected during those earlier campaigns while still predicting what volunteers might say now, we adapt our models to separately predict what a typical volunteer would have answered had they voted during each campaign.

GZ DESI includes fainter ($r < 19.0$ vs. $r \lessapprox 17.77$), smaller (see Sec. \ref{sec:apparent_vs_absolute}), and higher redshift ($z \lessapprox 0.4$ vs. $z < 0.15$) galaxies than Galaxy Zoo DECaLS. 
We will show that our predictions for the apparent morphology of these galaxies are reliable (Sec. \ref{sec:results}).
However, the difficulty in imaging morphology under such constraints causes apparent morphology to be an increasingly poor proxy for absolute morphology (i.e. the morphology that would be observed if the galaxy were closer).
Researchers using our catalogues should ensure their conclusions are not sensitive to this observational bias.

This paper is structured as follows.
We summarise the data available from the DESI-LS sister surveys (DECaLS, MZLS and BASS) and our approach to selecting galaxies and constructing images in Sec. \ref{sec:data}.
We describe training our deep learning models in Sec. \ref{sec:morphology_classifications}, focusing on our new approach to learn from multiple Galaxy Zoo campaigns simultaneously (unlike in GZ DECaLS itself). We measure and compare the accuracy of our models against other approaches in Sec. \ref{sec:results}. Finally, we apply the trained models to all three sister surveys and introduce our catalogues in Sec. \ref{sec:guidance}.\\

The morphology catalogues are available for download from \href{https://doi.org/10.5281/zenodo.7786416}{Zenodo}, CDS/Vizier, and NOIRLab's Astro Data Lab.
See Appendix \ref{appendix:data_availability} for further details. The code and weights for our deep learning models are available via \href{www.github.com/mwalmsley/zoobot}{GitHub}.

\section{Data}
\label{sec:data}

\subsection{Surveys}
\label{sec:surveys}

The Dark Energy Spectroscopic Instrument (DESI) is a cosmology-focused multi-object fibre spectrograph at the 4m Mayall telescope on Kitt Peak, USA.
DESI requires images to target its spectroscopic fibers; these are primarily provided by the DESI Legacy Surveys (DESI-LS).

DESI-LS is composed of three individual surveys working in concert; DECaLS, BASS, and MzLS.
BASS and MzLS cover the northern sky from Kitt Peak, USA; BASS captures  $g$ and $r$-band images using the Bok 2.3m telescope and MzLS captures $z$-band images with the same 4m Mayall telescope as DESI itself. We refer to both surveys jointly as BASS/MzLS.
DECaLS captures $grz$ images of the southern sky from the 4m Blanco telescope at Cerro Tololo Inter-American Observatory, Chile.
Together, DECaLS and BASS/MzLS provide 14,000 deg$^2$ of $grz$ targeting images for DESI.

Also noteable is the Dark Energy Survey (DES), an imaging survey focused on photometric redshifts. 
DES is not technically part of the DESI-LS; the primary survey footprint of $\delta < -18$ is too far south to be observed by DESI from Kitt Peak.
However, DES is being conducted with the same instrumentation as DECaLS (DECam on the 4m Blanco telescope), and so DES imaging is included in the DESI-LS data releases.

Specifically, DESI-LS DR8 includes all 5,000 deg$^2$ of $grz$ imaging taken by DECam as part of DES and released in DES DR2.

The four surveys (DECaLS, BASS/MzLS, and DES) together cover a combined area of 19,437 deg$^2$.
Their imaging properties are similar by design; DESI requires depths to be `as uniform as possible across the survey footprint' for consistent target selection \citep{Dey2018}.
This was successfully achieved.
The DESI-LS website\footnote{\href{https://www.legacysurvey.org}{https://www.legacysurvey.org}} shows median coadded depths (5$\sigma$ detection of a point source) of approximately $g=24.8$, $r=24.2$, and $z=23.4$ for DECaLS, and $g=24.2$, $r=23.8$, and $z=23.3$ for BASS/MzLS. DES DR2 quotes a median coadded catalogue depth for a 1\arcsec 95 diameter aperture at signal-to-noise ratio = 10 of
$g=24.7$, $r=24.4$, and $z=23.1$. 
This unique combination of deep and wide images is ideal for large-scale morphology classification.
Consistent imaging properties allow us to train deep learning models on volunteer classifications for a subset of images and then predict what volunteers would say for the remainder.

A subset of DECaLS images in DESI-LS DR5 were previously classified by Galaxy Zoo volunteers. These volunteer classifications were released as part of Galaxy Zoo DECaLS (W+22). The GZ DECaLS volunteer classifications provide the bulk of the training data we use in this work. We describe the GZ DECaLS subset selection and labelling process in more detail in Sec. \ref{sec:morphology_classifications}.

\begin{figure}
    \centering
    \includegraphics[width=\columnwidth]{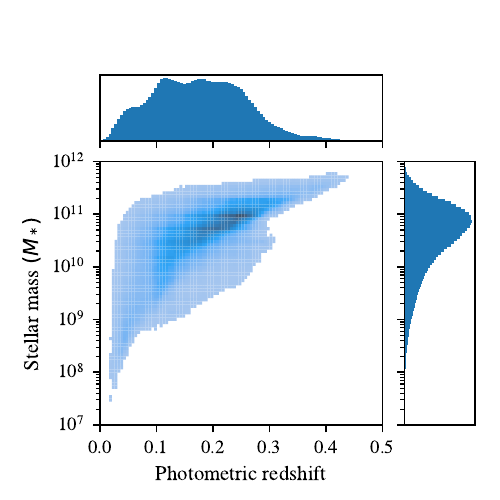}
    \caption{
    Photometric redshifts and estimated stellar masses of GZ DESI galaxies. Measurements from \citealt{Zou2019} (see Appendix \ref{sec:crossmatching}).
    }
    \label{fig:redshift_vs_mass}
\end{figure}

\subsection{Source Identification and Photometry}
\label{sec:source_identification}

The DESI-LS source database (i.e. the coordinates and basic photometry of identified sources in the DESI-LS images) is constructed using the Bayesian sourcefinding tool \texttt{tractor} \citep{tractor2016}. We query the \texttt{tractor} database using the Table Access Protocol (TAP) server made available by the DESI-LS collaboration at \href{https://datalab.noirlab.edu/tap}{https://datalab.noirlab.edu/tap} to select all extended\footnote{i.e. of \texttt{tractor} class other than `PSF'} sources with $r$ < 19.0. We selected this cut by visual inspection with the aim of identifying an approximate limit beyond which galaxies rarely have meaningful resolved visual morphology.

To remove stellar contamination, we exclude sources with an approximate surface brightness lower than 18 mag arcsec$^{-2}$, calculated as
\begin{equation}
    \mu = \textrm{mag}_{\rm r} + 2.5 \log_{10} \left( \pi r^{2}\right)\;,
\end{equation}

where $\mu$ is the surface brightness, $\textrm{mag}_{\rm r}$ the r-band magnitude and $r$ the estimated radius. The radius was estimated using:
\begin{equation}
    r = f \: r_{\rm DeV} + (1-f_{\rm Dev}) \: r_{\rm Exp}\;,
\end{equation}

where  $r_{\rm DeV}$, $f_{\rm Dev}$ and $r_{\rm Exp}$ are photometric properties estimated by \texttt{tractor}. In short, \texttt{tractor} models sources as a weighted mixture of an exponential (Exp) and a De Vaucouleurs (DeV) light profile. It reports the fraction of light attributable to each profile ($f_{\rm Dev}$, $f_{\rm Exp}$) and the angular half-light radius of those profiles by band (e.g. $r_{\rm DeV}$, $r_{\rm Exp}$).\\

Our magnitude $r$ < 19.0, surface brightness $\mu > 18$ mag arcsec$^{-2}$ and non-PSF selection leads to a total of 8,956,477 sources.
We download reduced flux measurements of each source using the DESI-LS cutout service. We refer to these measurements as native images. Native images are downloaded at telescope resolution (0.262 arcsec per pixel) with a field-of-view as similar as possible to the field-of-view which GZ DECaLS would have used, had the galaxy been included in GZ DECaLS. This is to make the new images on which we make predictions as similar as possible to the GZ DECaLS images used for training (i.e. to minimise the distribution shift). We describe the details of this field-of-view calculation in Appendix \ref{sec:field-of-view}.
The field-of-view is calculated according to the estimated radius of the galaxy, with the aim that galaxies of different angular size (due to e.g. greater distance) fill a consistent portion of the image.

\subsection{RGB Images}
\label{sec:rgb_images}

We next convert the native images to RGB images suitable for human and automated classification.
As with field-of-view, we minimise distribution shift by following the same process as GZ DECaLS.

Images are resampled from arbitrary pixel dimensions at native telescope angular resolution to 424x424 pixels at arbitrary angular resolution. 
We repeat the same colouring process as GZ DECaLS (W+22), which typically leads to images with less pronounced colour than in e.g. GZ2 \citep{Willett2013}.

Images with more than 20\% of flux measurements missing in any band are discarded. 
8,733,858 (97.8\% of 8,925,926) are successfully downloaded from the DESI-LS cutout service, of which 8,689,370 (99.5\%) have no more than 20\% missing flux in any band. Our final sample is thus these 8,689,370 galaxies.

\section{Morphology Classifications}
\label{sec:morphology_classifications}

Our goal is to provide accurate morphological classifications for every galaxy image in DESI-LS. 
It is not feasible to do this with volunteers alone.
GZ DECaLS collected 7.5 million individual volunteer classifications over 4.5 active years, for an average rate of approximately 1.7 million classifications per year. 
At that rate, collecting 40 classifications per DESI-LS galaxy (the standard prior to GZ DECaLS) would take approximately 200 years.
Collecting 5 classifications, the minimum used by GZ DECaLS (which prioritised volunteer effort towards galaxies most informative for training models, W+22), would still take an impractical 25 years.
We must therefore rely on automated methods for most galaxies.

We have two sources of volunteer classifications with which we can train models to make predictions on new DESI-LS images.
Our first source of labels is the 7.5 million GZ DECaLS classifications mentioned above.
A subset of these were already used for training GZ DECaLS models.
However, changes to the Galaxy Zoo website during the project mean classifications collected early in GZ DECaLS are not necessarily equivalent to classifications collected at the end.
Sec. \ref{sec:multicampaign_training} describes our method for resolving this shift to train on all GZ DECaLS classifications.
Our second source of labels is the additional volunteer classifications collected subsequently to GZ DECaLS (DESI-LS DR8).
We use the new classifications of DESI-LS DR8 as additional training data to improve our models, particularly in the faint, high-$z$, low-angular-size regime not covered by GZ DECaLS.
Sec. \ref{sec:additional_labels} describes the collection of our additional labels.

\subsection{Multi-Campaign Training}
\label{sec:multicampaign_training}

GZ DECaLS is our key source of training labels, providing 70\% of our 10M volunteer classifications (76\% of our 401k labelled galaxies).
Unfortunately, learning from these labels is complicated by the fact that labels collected at the start of GZ DECaLS are not equivalent to labels collected at the end.
This phenomenon is generally known as \textit{label shift} \citep{Storkey2013}.
Below, we describe the context and causes of our label shift and then discuss how we work around it to train models on all labels simultaneously.

\subsubsection{Context and Causes of Label Shift}
\label{sec:label_shift}

GZ DECaLS collected classifications over three Galaxy Zoo campaigns; GZD-1, GZD-2, and GZD-5, covering DECaLS images first released in DESI-LS DR1, DR2, and DR5.
GZD-1 and GZD-2 ask identical questions (we will often group them as GZD-1/2).
GZD-5 adjusted those questions based on preliminary results from GZD-1/2 and on the developing science interests of the community.
Some questions were unchanged, some had minor adjustments (e.g. from four to five possible bulge size answers) and some were entirely reworked (e.g. to improve sensitivity to weak bars and minor mergers). 
Volunteers were therefore asked different questions, with different possible answers, for GZD-1/2 vs. GZD-5 galaxies - a clear incompatibility for standard supervised training approaches.

Even where the questions and possible answers are unchanged, volunteers might systematically select from those possible answers in slightly different ways.
Questions were clarified though updates to the descriptive answer icons, the tutorial, and the field guide.
These had the explicit intention of slightly altering the distribution of answers that volunteers select, in order to better match the scientific aim of the question.
Unintended changes are also possible; over the seven years between the start of GZD-1 and the end of GZD-5, the population of volunteers itself may have gradually shifted.
These intentional and unintentional changes mean that volunteers selecting a given answer to a given question should not be interpreted in exactly the same way.

In short, GZD-1/2 and GZD-5 asked slightly different questions and received slightly different distributions of responses.
We therefore cannot naively use the responses for GZD-1/2 and GZD-5 as interchangeable training labels.

The models used to create the Galaxy Zoo DECaLS automated catalogue side-stepped this issue by training only with responses from the GZD-5 campaign.
This strategy worked but was not optimal.
Responses are roughly equally divided between GZD-1/2 (3.43M responses) and GZD-5 (3.84M responses) and so the GZ DECaLS models were trained using only half of the responses available.
Previous work shows that Galaxy Zoo models perform better where more training labels are available \citep{Walmsley2020} and so simplifying the training process by discarding half of the labels likely reduced model performance.
Further, volunteers kindly contribute their time to labelling galaxies and so we have a responsibility to use those labels efficiently i.e. to derive as much science value as possible.
We should ideally use all the training labels available. To do so, we introduce a new multi-campaign loss function that allows models to learn from several Galaxy Zoo campaigns simultaneously.

\subsubsection{Multi-Campaign Loss Function}
\label{sec:loss}

Different Galaxy Zoo campaigns asked different questions with different possible answers and received different distributions of volunteer responses.
We would like to learn from all of these responses across campaigns, in order to maximise our training data and create better models.
To do so, we will treat predicting the responses for each campaign as separate prediction tasks that use the same shared representation.
We provide the mathematical details below.

Consider a scenario where some Galaxy Zoo campaign A asks question $q_A$ for galaxies $G_A$ and campaign B asks question $q_B$ for galaxies $G_B$. 
Assume we can encode the volunteer responses (label) for some galaxy $g$ as vectors $k_A$ or $k_B$, depending on which campaign labelled the galaxy.
For example, campaign GZD-1/2 asked volunteers if a galaxy had a bar, with possible answers `Yes' or `No', while campaign GZD-5 offered possible answers `Strong', `Weak', or `No'. We could encode the volunteer response as vote counts e.g. $k_A=[3, 0]$ if 3 volunteers responded `Yes' during GZD-1/2 or as $k_B=[2, 1, 0]$ if 2 volunteers responded `Strong' and 1 volunteer responded `Weak' during GZD-5. Note that $k_A$ and $k_B$ may be different lengths.

One simple way to train a single model to predict answers to both campaigns would be to write a loss function that, 
if the galaxy $g \in G_A$, treats the model outputs as a prediction of $k_A$, and vice versa for $B$.
A straightforward implementation would be to concatenate $k_A$ and $k_B$ (where one would be filled with default/masked values), use a model with fixed output dimension $D_\text{model} = D(k_A) + D(k_B)$, and a loss function that ignores default/masked values (and hence provides gradients that depend only on the relevant question). 
Conveniently, the Dirichlet-Multinomial loss function introduced in W+22 is just such a function.

For each question, the loss takes the form:

\begin{equation}
    \label{multivariate_per_q_likelihood}
    \mathcal{L}_q = \int \text{Multi}(\vec{k}|\vec{\rho}, N) \text{Dirichlet}(\vec{\rho}| \vec{\alpha}) d\vec{\rho}
\end{equation}

where, for some target question $q$, $\vec{k}$ is the (vector) counts of responses (successes) of each answer, $N$ is the total number of responses (trials) to all answers, and $\vec{\rho}$ is the vector of probabilities of a volunteer giving each answer. $\vec{\rho}$ is drawn from $\text{Dirichlet}(\vec{\rho}|\vec{\alpha})$, where the model predicts the Dirichlet concentrations $\vec{\alpha}$. Intuitively, this loss corresponds to the odds of observing $k$ heads (votes for an answer) after $N$ coin flips (volunteers asked) assuming a (model-predicted) distribution for the bias of that coin. 
See W+22 for an extended description.

Assuming answers to different questions are independent, the loss may be applied to multiple questions via the sum
\begin{equation}
    \log \mathcal{L} = \sum_q \mathcal{L}_q(\vec{k_q}, N_q, \vec{f^w_q})
\end{equation}
where, for question $q$, $N_q$ is the total number of votes for all answers, $\vec{k_q}$ is the observed votes for each answer, and $\vec{f^w_q}$ is the predictions of our deep learning model for all answers (which we interpret as the Dirichlet $\vec{\alpha}$ parameters in Eqn. \ref{multivariate_per_q_likelihood}).

W+22 introduced this loss in the context of questions where \textit{some} answers may have 0 votes. Here, we consider the context where \textit{all} answers may have 0 votes (because the question is not asked in the campaign).
When all answers have 0 votes, $p(a=0|\vec{\alpha}, N=0)=1$ for all $\vec{\alpha}$ and hence $\frac{\partial \mathcal{L}}{\partial \vec{\alpha}} = 0$, meaning unanswered questions do not affect the training gradients. 
The loss naturally handles questions with no answers.
We can therefore train a single model to predict answers to different questions in different campaigns.

What about if the same question is asked in multiple campaigns, but volunteers give systematically different answers?
Such a scenario is likely here due to e.g. clarified instructions over the course of GZ DECaLS (see Sec. \ref{sec:label_shift}).
Our brute-force solution is to always consider questions as different between campaigns, even if the question itself has not changed.
We construct a multi-question multi-campaign label vector $\vec{K}$ where $K_i$ is the votes for answer $i$ and $i$ indexes all answers across all questions \textit{across all campaigns}. For a galaxy labelled in any single campaign, $K_i$ is 0 for any answer $a_i$ to any question not asked in that campaign. Every answer is always predicted but the prediction only affects training if votes for that answer are non-zero. Intuitively, this corresponds to having zero recorded votes to questions not asked in that campaign. Questions in different campaigns (GZD-1/2, GZD-5, and GZD-8) are effectively treated as separate prediction tasks using the same representation.  With this setup, the model learns a shared representation for predicting every answer to every question in every campaign, even when the questions are different or the distributions of answers have changed.

\subsection{New Volunteer Labels}
\label{sec:additional_labels}

Our multi-campaign loss allows us to jointly learn from volunteer responses to multiple Galaxy Zoo campaigns.
While the GZ DECaLS models were trained only on GZD-5 responses, we can now learn from both GZD-1/2 and GZD-5.
Further, we can also run new campaigns to collect new responses, and jointly learn from those new responses as well.

Following the conclusion of GZD-5 (November 2020), we asked Galaxy Zoo volunteers to label DECaLS images newly released in DESI-LS DR8.
We later expanded this to include all DESI-LS DR8 images (i.e. also including MzLS and BASS).
Galaxies were randomly selected from the catalogue described in Sec. \ref{sec:source_identification}, excluding galaxies already classified by volunteers in GZ DECaLS.
RGB images were constructed as described in Sec. \ref{sec:rgb_images}, except that the field-of-view was directly set by the weighted half-light radius $f_{\rm DeV}$ (see Sec. \ref{sec:source_identification}) rather than approximating the NSA Petrosian radii (see Appendix \ref{sec:field-of-view}) as work on those approximations was still ongoing at the time.
The classification procedure was identical to GZD-5 (i.e. we made no further changes to the Galaxy Zoo website itself).
We refer to the campaign gathering these new labels as GZD-8.

While the imaging quality is comparable to GZ DECaLS, the galaxies classified are dramatically different.
Recall that GZ DECaLS required galaxies to be listed in the NASA-Sloan Atlas (NSA).
The NSA was derived from SDSS images and hence typically has $r \lessapprox 17.77$ and $z < 0.15$.
Removing the NSA requirement removes this additional selection function and hence the newly-classified DR8 galaxies (and indeed all DR8 galaxies) are generally higher redshift, smaller, and fainter than those classified in GZ DECaLS.

Between Nov. 2020 and Oct. 2022, 38,949 volunteers\footnote{Unique cryptographic hashes derived from IP addresses are used to group classifications made by volunteers not logged in (13\%). The original IP addresses are never revealed to GZ researchers.} made 3.2M classifications of 105k galaxies. As with GZ DECaLS\footnote{See W+22 Sec. 4.3.1 for a discussion}, we remove as statistically unlikely the classifications of 347 users (0.9\%) who classified at least 150 galaxies and answered `artifact' for a majority of those galaxies.
Unlike earlier Galaxy Zoo works, but in keeping with Galaxy Zoo DECaLS, we do not attempt to re-weight volunteer votes to improve consistency (`weighted vote fractions').
We use the new responses to DESI-LS images as additional training data, which we anticipate will be particularly helpful for making accurate automated classifications of fainter, higher-redshift galaxies not previously labelled. 

\subsection{Model and Training Details}

Our model is a variant of EfficientNetB0 \citep{Tan2019a} with the classification head (i.e. the final layer following the global max pooling) replaced by an alternative head suitable for predicting Dirichlet concentrations.
Specifically, the final layer has 98 units, each with a sigmoid activation function scaling the outputs to fall between 1 and 101\footnote{Outputs below 1 allow for bimodal vote fraction posteriors, which would complicate our reporting of the mean vote fractions and associated uncertainties. The 1-101 range sets the maximum posterior skew towards a given answer, which we find useful for numerical stability when training with mixed precision (at the cost of restricting expressivity for the most extreme galaxies)}.
We chose EfficientNetB0 as our base architecture to balance performance with practicality.
The EfficientNet family is designed to achieve high accuracy relative to their fast (here, 15ms/galaxy) inference speed. B0 is also small enough (approx. 4M parameters) to be trainable on consumer-grade GPUs, which we consider critical to making our models useful for other astronomers.
Appendix \ref{sec:training_details} provides full details of constructing our training sets, defining and training our models, and selecting our hyperparameters.

\section{Model Performance Results}
\label{sec:results}

The goal of this paper is to accurately measure visual morphology for every well-resolved galaxy in DESI-LS (Sec. \ref{sec:data}).
The scale of DESI-LS requires that we first train models to reproduce the responses of Galaxy Zoo volunteers for a small subset of DESI-LS galaxies, and then use those models to predict how GZ volunteers would respond for the bulk of the galaxies.
We therefore need to carefully check that our models do indeed accurately reproduce the GZ volunteer responses.
GZD-8 is a random labelled subset of the DESI-LS galaxies and so we will measure, and aim to maximise, performance at reproducing GZD-8 responses.

In Sec. \ref{sec:performance_vs_single}, we show how training on all campaigns improves performance at reproducing responses to every GZD-8 question.
In Sec. \ref{sec:breakdown}, we provide additional metrics for the GZD-8 performance of our best-performing (trained on all campaigns) model design.
In Sec. \ref{sec:literature}, we compare our predictions and performance metrics to other automated morphology catalogues in the literature.

\subsection{Performance vs. Single Survey Training}
\label{sec:performance_vs_single}

The loss we introduce (Sec. \ref{sec:loss}) allows us to simultaneously train to predict volunteer votes for several campaigns. Here, we show that training on all campaigns at once significantly improves performance at every GZD-8 question.

We have three campaigns to learn from: GZD-1/2, GZD-5, and GZD-8. We want to maximise performance on GZD-8, being a random subset of our DESI-LS catalogue. Which campaigns should we choose to learn on to maximise GZD-8 performance?

We investigate four training scenarios: GZD-8 only, GZD-5 only, GZD-1/2 and GZD-5, and all campaigns (GZD-1/2/5/8). We train five models for each training scenario, each on different random 70\%/10\%/20\% train/validation/test splits of the campaigns used. We report performance on that model's GZD-8 test set\footnote{If GZD-8 is not used for training, we simply pick a random 20\% of GZD-8 for evaluation.}.
Reporting the test set performance is valid because we make no further changes to the design; the model hyperparameters were previously chosen to maximise performance when training on GZD-5 only (App. \ref{sec:training_details}).

\begin{figure*}
    \centering
    \includegraphics[trim={0.7cm 0 0 0},clip]{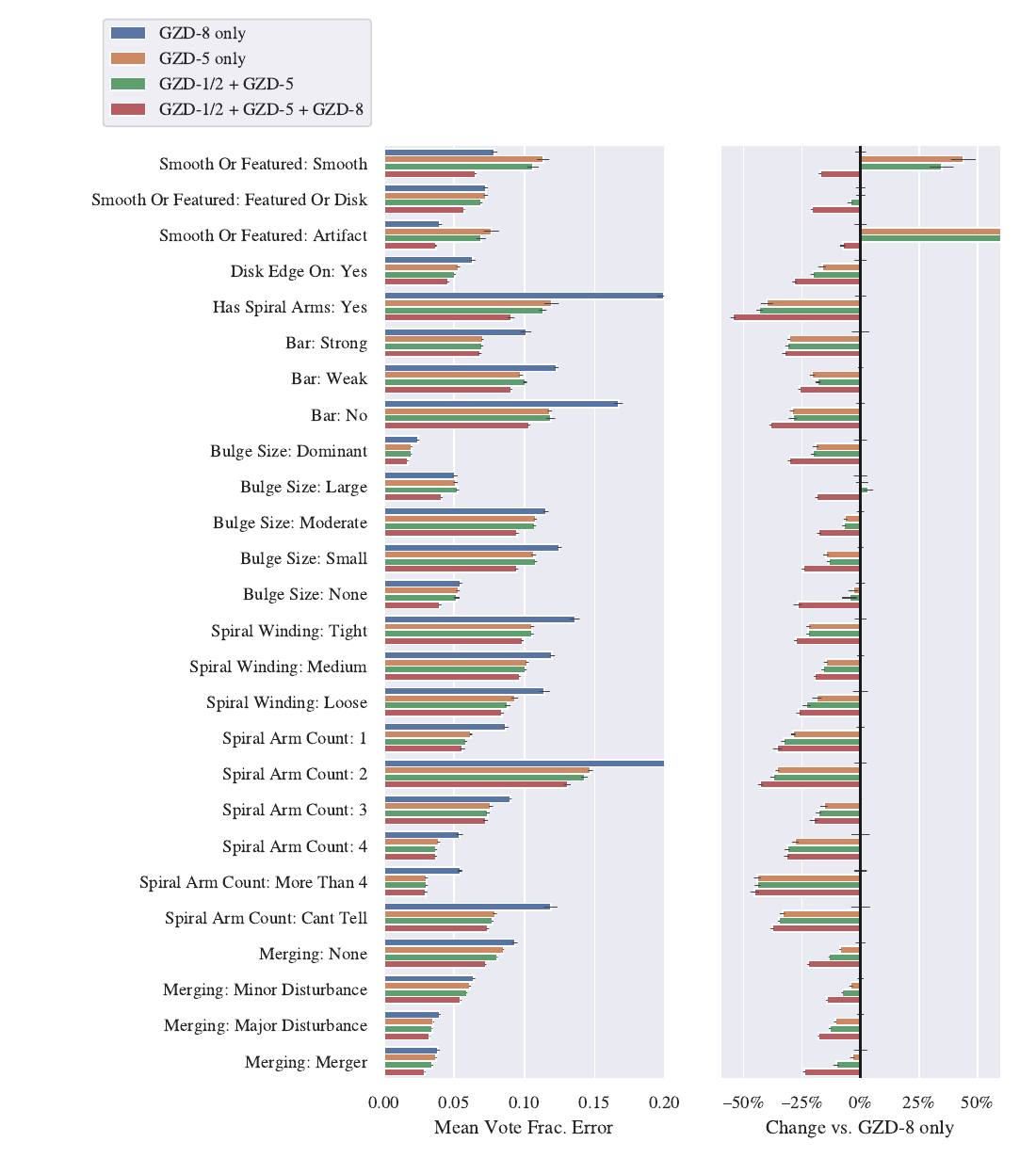}
    \caption{
    Training only on GZD-8 (blue, 96k galaxies) is typically worse than training only on GZD-5 (orange, 223k galaxies), likely due to GZD-8 being a smaller training sample. Adding GZD-1 and GZD-2 to GZD-5 (green, 305k) improves performance. Training on all four datasets (GZD-1/2/5/8, red, 401k) performs best at every question.
    \\
    \\
    Left: mean error when predicting GZD-8 volunteer vote fractions, for otherwise-equivalent models trained on different label subsets. 
    \\
    Right: as left, but showing \textit{percentage change} in mean vote fraction error vs. training only on GZD-8.
    \\
    \\
    Mean vote fraction error is the difference between the actual and predicted fraction of volunteers giving a specific answer, averaged over all GZD-8 test galaxies. Errorbars show the 95\% confidence interval calculated from five trials with different seeds.
    }
    \label{fig:mse_vs_campaign}
\end{figure*}

Fig. \ref{fig:mse_vs_campaign} shows model performance for each training scenario, broken down by question and answer. We report the mean vote fraction error (i.e. the difference between the actual and predicted fraction of volunteers giving a specific answer, averaged over all GZD-8 galaxies) for each training scenario.

We find that \textbf{training on all campaigns and predicting on GZD-8 performs best} and dramatically outperforms training on GZD-8 alone. The vote fraction error improves for every question. Typical improvements are in the range 20\%-40\%, with an average improvement over all questions of 27\%. We discuss each comparison in detail below.

Our baseline scenario is training only on GZD-8. This follows the typical approach in the literature of training and testing on subsets of the same labelled galaxy catalogue. Using random subsets of the same catalogue avoids issues including label shift described above (Sec. \ref{sec:label_shift}).
However, the GZD-8 trained models underperform our second scenario: training only on GZD-5. 

Training only on GZD-5 represents taking the models trained for GZ DECaLS, which only learned from GZD-5, and directly predicting what GZD-8 volunteers would say for DESI-LS galaxies. Without GZD-8 responses to train on, the model cannot predict GZD-8 responses themselves. Instead, we take the predicted GZD-5 responses and compare them to the actual GZD-8 test set responses. 
We find that the models outperform models trained only on GZD-8 (above) at all questions other than `smooth or featured?'. 
We believe this is because there are far fewer labelled galaxies in GZD-8 (96k) than in GZD-5 training responses (223k) and so any label shift between GZD-5 and GZD-8 is outweighed by the change in training dataset size.
Additional experiments (not shown) training on an equal number of GZD-1/2 and GZD-5 galaxies (96k) confirm that training on fewer galaxies reduces performance, as expected.\footnote{Interestingly, training on 96k GZD-1/2/5 galaxies outperforms training on 96k GZD-8 galaxies (not shown). This may suggest that the active learning system used during GZD-5 (W+22) successfully prioritised labelling the galaxies most informative for training models, or may simply reflect that the lower redshift of GZD-1/2/5 galaxies makes the typical GZD-1/2/5 galaxy more informative than GZD-8 galaxies.}

Next, we use our multi-campaign loss to train on both GZD-1/2 and GZD-5.
Introducing GZD-1/2 improves model performance for all questions over the GZD-5 trained version - even though we have still not introduced any GZD-8 training labels.
This suggests that learning to predict GZD-1/2 responses helps the model predict GZD-5 responses, which in turn better match the GZD-8 test responses. 

Finally, we can additionally train on the newly-collected GZD-8 labels to further improve performance.
Models trained on all campaigns - GZD-1/2, GZD-5, and GZD-8 - perform best.
The models can learn an effective representation from the large combined training set (10.3M total responses) while also benefiting from training labels specifically drawn from the same distribution as the test set.

\subsection{Performance Breakdown}
\label{sec:breakdown}

We described above how our multi-campaign loss improves model performance when compared to alternative approaches. We now provide further practical details on the final performance of our models. 

Appendix \ref{sec:confusion_appendix} shows confusion matrices when predicting the majority volunteer response for GZD-8 galaxies. 
When calculating confusion matrices for each question, we only consider galaxies where a majority of volunteers were asked that question. This is because not every question is relevant to every galaxy. It is not meaningful to measure whether we can accurately predict whether volunteers would label smooth galaxies as having tight or loose spiral arms, for example. Filtering to galaxies where at least half the volunteers were asked that question ensures that the predictions are relevant to each galaxy\footnote{We recommend this as a general practice and mark relevant galaxies/questions in our catalogues. See the Zenodo repository for instructions.}.

Confusion matrices for the majority response are useful for building an intuition for typical performance, but dividing galaxies into discrete classes (e.g. `Smooth or featured?') neglects the rich information our model provides (e.g. `how smooth?', as measured by the predicted `smooth' vote fraction).
We recommend that practicing astronomers use our posteriors or predicted vote fractions (see e.g. \citealt{Smethurst2015}) rather than binning to the most likely response.

How can we measure the accuracy of our predicted vote fractions?
If we asked an infinite number of volunteers then we could straightforwardly compare the predicted and true vote fractions.
However, we don't know what an infinite number of volunteers would say - we only know the observed vote fractions from our finite (N=1--40) crowd of volunteers.
There is an intrinsic counting uncertainty around the true vote fractions.
To work around this, we calculate analytic posteriors for the true vote fractions given our observed volunteer votes\footnote{The posterior for the bias of a coin assuming a uniform prior is well-known to be $p(\theta|k, n) = \beta(k+1, n-k+1$)}.
 We then sample possible observed vote fractions and measure their typical deviation from the now-known true vote fractions.
In short, we are constructing a realistic set of possible true vote fractions and then simulating the error introduced by trying to measure observed vote fractions with $N$ volunteers.
This is the minimum error that even a perfect model would experience.

\begin{figure}
    \centering
    \includegraphics{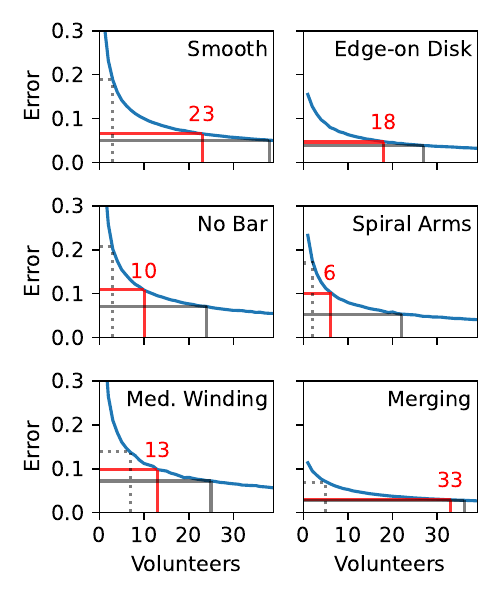}
    \caption{Vote fraction errors observed for our model (red) compared against the minimum statistical errors introduced by asking a finite number of volunteers (grey solid). A naive baseline (always guessing the mean vote fraction) is also shown (grey dotted). The blue curve shows the expected statistical error from asking $N$ simulated volunteers, where $N$ may be higher or lower than the actual number of volunteers asked. We show errors for a selection of key questions and answers.}
    \label{fig:deviation_vs_volunteers}
\end{figure}

Fig \ref{fig:deviation_vs_volunteers} compares the vote fraction errors measured by our model (predicted vs. observed) against the minimum possible error (true vs. observed in our catalogue).
The minimum error is attributable to statistical chance and any additional error is introduced by our model.
We also show a baseline naive model which always guesses the mean vote fraction for that answer.
Finally, to build intuition, we also show the expected error from asking $N$ volunteers where $N$ may be more or less than the actual number of volunteers asked in our catalogue.
 
The model predictions are typically as accurate as asking that question to around 15 volunteers. 
On questions where the model performs best, like `smooth or featured?' and `edge-on disk?', the model is as accurate as asking approx. 20 volunteers.
On questions where the model performs relatively worse, like `spiral arms?' and `bar?', the model is as accurate as asking 6-10 volunteers.
For questions with more than two answers, like `merger?' (with answers no/minor disturbance/major disturbance/merging), performance varies by answer similarly to Fig. \ref{fig:mse_vs_campaign}.
The model approaches the theoretical maximum agreement with our observed vote fractions for the edge-on disk and merger questions.

\subsection{Comparison to Other Automated Catalogues}
\label{sec:literature}

We would like to compare our classifications to the results of other automated approaches. There are no other automated catalogues with comparable imagery making measurements of comparable detail. However, several recent works have used deep learning to measure simpler visual morphology in DES imaging. DES imaging is a subset (38\% by area) of the DESI imaging classified in this work and therefore some galaxies are classified in both this work and other works.
In this section, we compare our measurements with those DES-focused works.

\subsubsection{What to Compare}

 Several authors (introduced below) have used deep learning, trained on Galaxy Zoo datasets, to predict visual morphologies for DES galaxies. However, these authors used different volunteer labels (from GZ1 or GZ2) which in turn were created with different images (SDSS). 

We therefore make comparisons by using our trained networks to predict the responses that would be obtained by GZ1 and GZ2 volunteers shown SDSS images. This is a similar but distinct task to predicting responses from GZ-DESI volunteers who have seen DESI images. While the predicted GZ1/GZ2 labels will not be as useful for most morphological science, as the deeper DESI images make the observed morphology more likely to reflect the `true' intrinsic morphology (Sec. \ref{sec:apparent_vs_absolute}), they do allow us to compare our approach with others. 

We make our comparisons using ROC AUC score (receiver operator characteristic area-under-the-curve score, \citealt{Fawcett2006}).
Our scenario is that we have several models, each trained on previous prediction tasks and outputting a scalar score, and we would like to measure performance on a new but similar prediction task.
For example, we would like to measure how the DESI-LS models (trained on DESI-LS) perform at predicting GZ1 responses, in order to compare performance on GZ1 with prior work that predicts GZ1 responses.
Since the new task is similar to the tasks for which each model was trained, the model scores should be reasonably correlated with the labels for the new task.
However, they will not (in general) be calibrated; we have no reason to expect a score threshold of 0.5 to be the ideal dividing line between a prediction of class 0 or class 1.
The ROC AUC score avoids the problem of calibration by summarising the true positive and false positive rates at every possible score threshold.
The ROC AUC score is therefore a convenient measure for how well the scores from a model trained on a previous task correspond to the labels of the new prediction task.

We also consider the scenario of `how well could we do at predicting the new task, if we didn't have any machine learning models but we did have volunteer answers for every galaxy from an earlier GZ campaign'.
We calculate ROC AUC scores using the volunteer vote fractions for the previous task as prediction scores for the new task, as if the volunteers were models making predictions.
For example, Table \ref{tab:ml_metrics} shows the ROC AUC scores calculated using the fraction of DESI-LS volunteers who answered `smooth' as a score and the fraction of GZ1 volunteers who answered `elliptical' as a label.

Using the published predictions of other authors introduces a relative disadvantage for our DESI-LS models. 
Models generally have higher measured performance for the galaxies on which they were trained (the training set) than on unseen comparable galaxies (e.g. the test set). 
In practice, performance on the test set is the appropriate measurement; we want to make predictions on new galaxies, not galaxies for which we already have labels.
We know the galaxies on which our DESI-LS models are never trained (the test set), and so we can ensure we only measure performance on these unseen galaxies.
However, for the models of other authors, we do not know on which galaxies the other models were trained. We have only the all-galaxy public prediction catalogues, and so we must measure performance on these. 
We therefore expect our measured performance for the models of other authors to fall in-between the (appropriate) lower test set performance and (misleadingly high) training set performance.
Our performance measurements reported here will therefore be slightly biased towards the models of other authors.
Despite this, we will ultimately show below that our model compares favourably.

\subsubsection{Cheng 2020/21}
\citealt{Cheng2020} trained a convolutional neural network (CNN) to predict Galaxy Zoo 1 (GZ1) volunteer labels given DES images. The authors identify $\approx$2850 galaxies which are imaged in DES Y1 Gold and labelled by GZ1 volunteers (using earlier SDSS images) as `confidently' (defined as greater than 80\% volunteer agreement) being either elliptical or spiral. They use these confident responses as class labels (e.g. 0 if confidently elliptical, 1 if confidently spiral) and train a CNN to predict the class labels from processed versions of the DES Y1 images.

\citealt{Cheng2021} (hereafter C+21) continued this work by repeating a similar process to train an ensemble of CNNs, and then making predictions of `elliptical' or `spiral' on DES Y3 images (21M). These predictions are released as a public catalogue and allow the comparisons made here.

To make our comparisons, we identify the subset of galaxies with elliptical/spiral predictions from C+21, DESI-LS test predictions from our models, and responses from both DESI-LS volunteers and GZ1 volunteers (for ground truth labels). 
We identify these 1618 galaxies by cross-matching the coordinates of galaxies with C+21 predictions and our DESI-LS galaxy catalogue (Sec. \ref{sec:data}).
We previously (Sec. \ref{sec:performance_vs_single}) trained models on train/validation/test subsets of DESI-LS and measured the quality of their test predictions.
We re-use those test predictions here, selecting the 20\% of galaxies with test predictions from a random DESI-LS model.

As scores, we use the predicted elliptical probability from C+21's models and the predicted `smooth' vote fraction from our models.
We then measure the ROC curves achieved by those scores against two definitions of target label.

\subsubsection{Dominguez-Sanchez 2018/22 and Vega-Ferraro 2021}

\citealt{Sanchez2018} (DS+18) trained models to predict (in addition to non-GZ tasks) how the majority of GZ2 volunteers would respond to the questions `smooth or featured?', `edge-on?, `bar?', `bulge prominence?, `edge-on bulge shape?', and `merger?' for galaxies imaged by SDSS DR7. 
They then publicly released predictions of GZ2 majority responses for the 670k galaxy subset of SDSS DR7 analysed by \cite{Meert2015} (approximately twice the size of GZ2 itself).
DS+18's general approach has since been used in several related later works on smaller scales for the MaNGA survey.
\cite{Fischer2019} used DS+18's methodology to predict the morphology of galaxies in the MaNGA survey (DR15, \citealt{Wake2017}) and shares those new predictions as a value-added catalogue.
\cite{Sanchez2022} extended this value-added catalogue to a later MaNGA data release (DR17) with iterative improvements (specifically, making adjustments to the representation of non-GZ labels and adding model ensembling). 

DS+18's public catalogue was used as training labels by \citealt{VegaFerrero2021} (hereafter VF+21) when constructing a new catalogue of predicted DES morphologies.
VF+21 released deep learning predictions for (in addition to non-GZ tasks) whether DES DR1 galaxies are edge or face-on, as measured by GZ2 volunteers.
The authors aimed to infer intrinsic morphology i.e. to predict what GZ2 volunteers would have said had the galaxy been observed at lower redshift.
In contrast, this work aims to infer apparent morphology.
We therefore compare this work to VF+21 only on galaxies which VF+21 would have used as low-redshift examples.

To make our comparisons, and similarly to the process above with C+21, we identify the subset of galaxies with labels from both GZ2 and DESI-LS, predictions from either DS+18 or VF+21, and test predictions from a random DESI-LS model.
These galaxies would be considered as low redshift (`$z_0$') examples by VF+21.
We first assess how well the models predict if galaxies are featured according to GZ2 or DESI-LS volunteers.
As scores, we use the predicted `featured' probability from DS+18's models and the predicted `featured' vote fraction from our models.
Next, we assess how well each model predicts if galaxies are edge-on according to GZ2 or DESI-LS volunteers.
We choose as scores the predicted `edge-on disk' probability from DS+18's models or VF+21's models and the predicted `edge-on disk' vote fraction from our models.
Because edge-on is only a relevant question if a galaxy is featured, we only assess performance on galaxies labelled by the majority of volunteers as `featured'.

\subsubsection{Results}

Our DESI-LS models match or outperform previous works at the tasks they defined (majority response to GZ1 `elliptical', GZ2 `smooth' and GZ2 `edge-on') without any training on GZ1 or GZ2 labels. 
Further, we can subsequently train our DESI-LS models on GZ1/GZ2 labels (finetuning, see \citealt{Walmsley2022Towards}) and outperform previous works in all cases.

Our figure-of-merit is the ROC `area under the curve' (AUC) score.
This can be interpreted as the probability that any two galaxies will be ranked in the correct order (according to the task labels) by the model scores.
We also show the highest accuracy achieved for any choice of score threshold (Acc.) to provide further intuition.

Table \ref{tab:ml_metrics_upper} shows performance metrics for each GZ1 or GZ2 task (i.e. predicting what a GZ1/2 volunteer would say).
We compare performance between the model of the author who selected that task, GZ DESI volunteers answering an equivalent question, and our DESI-LS models before and after finetuning.
When we finetune our DESI-LS models on GZ1 or GZ2 labels, and test performance on still-unseen GZ1/GZ2 labels, the finetuned models outperform previous works at every task.

Table \ref{tab:ml_metrics_upper} shows performance metrics for the equivalent DESI-LS tasks (i.e. predicting what a GZ volunteer would say).
Our models significantly outperform previous works at predicting volunteer responses for every equivalent DESI-LS tasks.
This is expected given our new DESI-LS labels but is important to note for astronomers looking for the most accurate morphology measurements on DESI-LS images.
\begin{table}

    \begin{subtable}[h]{\columnwidth}
        \centering
        
        \begin{tabular*}{\textwidth}{c @{\extracolsep{\fill}}lllrr}
        \toprule
         Task & Dataset   &                   Method &   AUC &   Acc. \\
        \midrule
        GZ1 Elliptical & C+21 $\cap$ GZ1 &                 C+21 & 0.858 & 78.1\% \\
                    & &             DESI Vols & 0.860 & 77.4\% \\
                    & &               DESI ML & 0.877 & 81.1\% \\
                    & &  \textbf{DESI ML + GZ1} & \textbf{0.955} & \textbf{88.9\%} \\
        \midrule
        GZ2 Smooth & DS+18 $\cap$ GZ2 &               DS+18 & 0.946 & 90.1\% \\
                    & &             DESI Vols & 0.935 & 88.1\% \\
                    & &               DESI ML & 0.967 & 91.3\% \\
                    &  &  \textbf{DESI ML + GZ2} & \textbf{0.974} & \textbf{92.4\%} \\
        \midrule
        GZ2 Edge-On & DS+18 $\cap$ GZ2 &               DS+18 & 0.992 & 97.0\% \\
                    & &             DESI Vols & 0.981 & 97.5\% \\
                    & &               DESI ML & \textbf{0.999} & 98.7\% \\
                    & &  \textbf{DESI ML + GZ2} & \textbf{0.999} & \textbf{98.9\%} \\
        \midrule 
        GZ2 Edge-On & VF+21 $\cap$ GZ2 &            VF+21 & 0.982 & 95.0\% \\
                    &                  & DESI Vols        & 0.997 & 98.5\% \\
                    &                  & \textbf{DESI ML} & \textbf{0.999} & \textbf{98.8\%} \\
        
        \bottomrule
        \end{tabular*}
        \caption
        {Performance metrics on tasks pursued by previous authors (Cheng+21, Sanchez+18). 
        Finetuning the final layer (`+GZ1', `+GZ2') of our DESI-LS models improves performance.
        }
        \label{tab:ml_metrics_upper}
     \end{subtable}
    \newline
    \vspace*{0.25 cm}
    \newline
    \begin{subtable}[h]{\columnwidth}
        \centering
        
        \begin{tabular*}{\textwidth}{c @{\extracolsep{\fill}}lllrr}
        \toprule
         Task & Dataset  &   Method &   AUC &  Acc. \\
        \midrule
        DESI Smooth & C+21 $\cap$ GZ1 &    C+21 & 0.712 & 68.7\% \\
                        &  &    GZ1 Vols & 0.829 & 76.4\% \\
                        &  &  \textbf{DESI ML} & \textbf{0.964} & \textbf{89.6\%} \\
        \midrule
        DESI Smooth     & DS+18 $\cap$ GZ2 &  DS+18 & 0.899 & 83.0\% \\
                        &  &    GZ2 Vols & 0.924 & 87.2\% \\
                        &  &  \textbf{DESI ML} & \textbf{0.955} & \textbf{88.9\%} \\
        \midrule
        DESI Edge-On & DS+18 $\cap$ GZ2 &  DS+18 & 0.984 & 96.1\%\\
                        & &    GZ2 Vols & 0.993 & 97.8\% \\
                        & &  \textbf{DESI ML} & \textbf{0.994} & \textbf{97.9\%} \\
        \midrule        
        DESI Edge-On & VF+21 $\cap$ GZ2 & VF+21 & 0.978 & 94.4\% \\
                     &                    &  GZ2 Vols & 0.996 & 98.1\% \\
                     &                    &  \textbf{DESI ML} & \textbf{0.998} & \textbf{98.8\%} \\

        \bottomrule
        \end{tabular*}
        
       \caption{Performance metrics on the equivalent DESI-LS tasks to those pursued by previous authors (above) using earlier (GZ1/GZ2) data.}
       \label{tab:ml_metrics_lower}
    \end{subtable}
      \caption{Performance metrics for the GZ1 and GZ2 tasks pursued by previous authors (above) and the equivalent DESI-LS tasks (below). Our models outperform or match the published predictions of previous authors at the tasks they pursued, and significantly outperform those published predictions at the equivalent DESI-LS tasks.}
     \label{tab:ml_metrics}
\end{table}

\section{Catalogues}
\label{sec:guidance}

\subsection{Automated Morphology Measurements}
\label{sec:prediction_details}

We train an ensemble of five models to make our final DESI-LS predictions.
We aim for our models to perform as well as possible and therefore provide them with as many labelled DESI-LS galaxies as possible.
We do not reserve a test set - performance measurements made on a test set are reported separately in Sec. \ref{sec:results}. 
We still reserve a small validation set (5\%) to use early stopping to avoid overfitting.
We use a different random 5\% for each model to diversify our early stopping condition.

Each model makes five predictions, each with dropout applied to the penultimate layer during the forward pass (a.k.a. MC Dropout, \citealt{Gal2016Uncertainty}).
Training each model takes approximately 10 hours using a pair of 40GB NVIDIA A100 GPUs.
Making five predictions (on a single A100) takes approx. 15ms/galaxy (4k/minute).
Distributed predictions for all 8.7M galaxies take a matter of hours on an institutional GPU cluster.
This is significantly quicker than the time taken to download cutouts from the DESI-LS cutout service (of order several weeks).

The final automated morphology measurements are available via \href{https://zenodo.org/record/7786417#.ZCcrqPdw3u0}{Zenodo}. 

\subsection{Volunteer Response Catalogue}

We anticipate that most astronomers will primarily use the automated morphology measurements.
These align well with the volunteer responses (see Sec. \ref{sec:results}) and cover dramatically more galaxies.
However, for astronomers specifically interested in volunteer responses, we also release our new GZD-8 volunteer votes (Sec. \ref{sec:additional_labels}).

\begin{figure}
    \centering
    \includegraphics{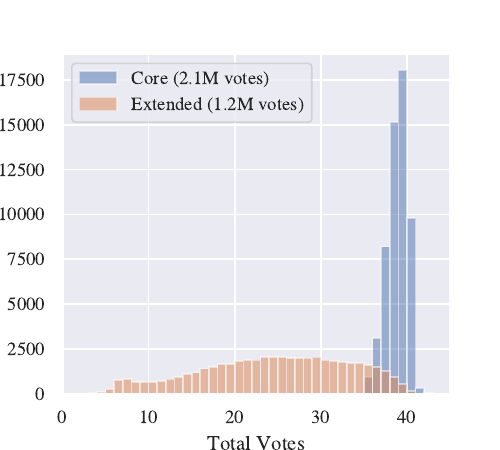}
    \caption{Total volunteer vote counts (after reliability filtering) for the `Core' and `Extended' samples of galaxies newly-labelled in GZD-8.}
    \label{fig:core_vs_extended}
\end{figure}

For clarity, we divide the newly-labelled galaxies into `Core' and `Extended' samples. 
The `Core' sample includes galaxies with at least 36 reliable (Sec \ref{sec:additional_labels}) volunteer votes and no more than five votes for `artifact'. 
Galaxies receiving more than five votes for `artifact' were immediately retired and receive no further votes.
These galaxies form the vast majority of the `Extended' sample.
`Extended' additionally includes a very small fraction of galaxies (1.7\%) which did not reach 36 total reliable votes. This is due either to being uploaded but not fully classified at the conclusion of DESI-LS, or to the vote reliability process (Sec \ref{sec:additional_labels}).
Figure \ref{fig:core_vs_extended} shows the total votes per galaxy in each sample.

We advise astronomers who require reliable volunteer vote fractions to begin with the `Core' sample only.
Astronomers aiming to find anomalies may be particularly interested in the `Extended' sample.
For training machine learning models, galaxies in the `Extended' sample have fewer votes and so aggregated class labels (e.g. majority vote fractions) will be noisier.
The models trained in this work use all DR8 votes (i.e. `Core' plus `Extended') as our loss function (Sec \ref{sec:loss}) uses the vote counts directly and so noisy vote fractions are not an issue.

\subsection{Apparent vs. Absolute Morphology}
\label{sec:apparent_vs_absolute}

The same galaxy will typically appear less featured when placed at higher redshift \citep{Bamford2009}. 
It is therefore crucial to separately refer to visible morphology (i.e. how a galaxy appears) and intrinsic morphology (which we define as how it would appear at some reference distance).
In analogy to magnitudes, we will refer to these as apparent and absolute morphology.
The catalogues presented here measure apparent morphology; how the galaxies appeared to Galaxy Zoo volunteers and to our models.

\begin{figure*}
    \centering
    \includegraphics{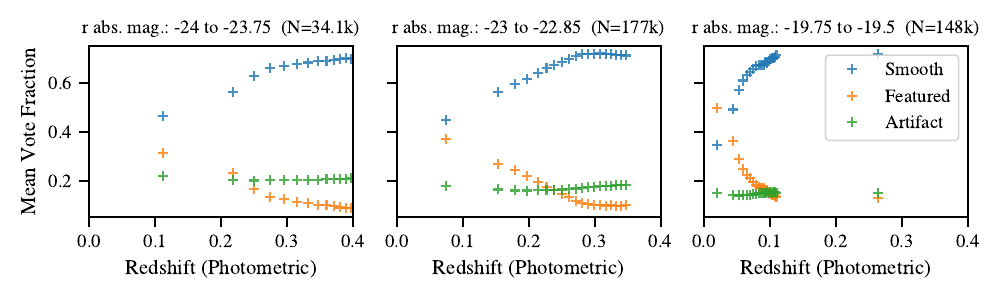}
    \includegraphics{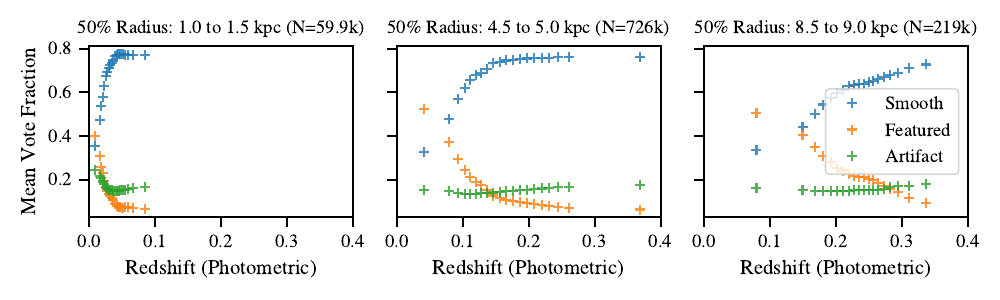}
    \caption{
    Trend with redshift of automated vote fractions for `Smooth or Featured'.
    Featured likelihood drops significantly from $z=0$ for all galaxy subsets.
    Split by radius, the largest galaxies ($\approx$9kpc) have a similar likelihood of featured votes to typical ($\approx$5kpc) galaxies but have a slower drop in featured likelihood as redshift increases.
    Markers indicate vote fractions in 20 equal-count photo-z bins. Statistical errors are negligible throughout.}
    \label{fig:frac_vs_redshift}
\end{figure*}

We expect brightness and radius to influence the change in morphology with increasing redshift. At high redshift, fainter features will become too dim to distinguish and smaller features will become too small to resolve.
Fig. \ref{fig:frac_vs_redshift} shows how galaxy subsets of different absolute magnitude and physical radius experience different changes in apparent morphology with redshift.

Split by absolute magnitude, fainter galaxies experience a more rapid decline in featured fraction than brighter galaxies. This matches our expectation that fainter galaxies will typically have fainter features and that these features will more rapidly fall below the limiting magnitude of our instrument.
Split by physical radius (specifically the Petrosian half-light radius), smaller galaxies similarly experience a more rapid decline in featured fraction than larger galaxies. This also matches our expectation that smaller galaxies will typically have smaller features and that these will more rapidly fall below the effective point-spread function of our instrument. 

\section{Discussion}
\label{sec:discussion}

This data release uses imaging from DESI-LS DR8. Subsequently to this work, additional imaging (particularly from DES in the southern sky) has been released in DESI-LS DR9 and DR10. Our models could be used to seamlessly extend our classification catalogue to include these new images. In general, we hope to ultimately shift to a pattern where labels are collected and models trained on current data releases, and then those models are immediately applied to new data releases as they become available.

We showed in Sec. \ref{sec:multicampaign_training} that training on GZD-5 and predicting on GZD-8 worked better than training directly on GZD-8, perhaps because the larger size of GZD-5 outweighs any label shift between GZD-5 and GZD-8.
The practical implication is that without the multi-campaign loss we introduce (\ref{sec:loss}),
we could have better measured the morphology of DESI-LS galaxies using existing GZD-5-trained models (W+22) than with a new labelling campaign and new models.
However, with our multi-campaign loss, we can learn from both the previous GZD-5 labels and the new GZD-8 labels and ultimately make better DESI-LS automated measurements than with either alone.
This removes a major trade-off for deciding when a new labelling campaign is valuable. 
Before, one could either train on existing plentiful labels gathered with different images or different questions, or collect smaller numbers of new labels for your images and questions.
Now, one can do both.

We noted in Sec. \ref{sec:prediction_details} that the time to make predictions on every image (of order hours when distributed on our GPU cluster) was far shorter than the time to download those images from the DESI-LS cutout service servers (of order weeks).
We anticipate that this will be a general trend for future large surveys; data transfer time will be the limiting factor rather than prediction time.
Remote access initiatives including LINCC and ESA Datalabs make it straightforward to run algorithms locally to the data, eliminating the need to transfer data.
This will be ideal for running inference with our final science-ready models.
But developing the models in this work took several thousand GPU-hours - small in the scale of recent computer science work, but substantially more than could reasonably be expected to be made available to all external astronomy researchers by remote access initiatives.
Unlike classical algorithms, developing machine learning models on a local data subset is not ideal, especially for algorithms which include unsupervised components (as the next generation of morphology classifiers will likely do).
How can we have both all of the useful data and major compute resources?
One path forward may be to identify data subsets that are useful to many practitioners (e.g. cutouts of all of the well-resolved galaxies) and transfer them to a shared HPC resource.
We urgently invite community planning on this topic.

This is the first Galaxy Zoo catalogue to include some galaxies measured only by volunteer-trained algorithms.
For GZ DECaLS, galaxies selected as likely to be most informative in training our models were prioritised for intensive volunteer labelling (40 votes) but all other galaxies were still inspected by volunteers (5 votes). 
For GZ DESI, covering the full DESI-LS footprint, even 5 votes was no longer feasible. 
We anticipate that potential future Galaxy Zoo catalogues for Euclid and Rubin will likely require classification at even larger scales and therefore include a greater fraction of algorithm-only morphology measurements.
The question of how to best allocate our limited volunteer votes will therefore be increasingly important.

More broadly, effective algorithmic systems may let Galaxy Zoo volunteers may take on increasingly diverse roles.
Volunteers may help us create small labelled datasets for a panoply of targeted science questions (e.g. rings, ram-pressure-stripped `jellyfish' galaxies, interacting galaxies, etc.) that label-efficient models then scale up to arbitrarily large surveys.
Volunteers may integrate `AI assistants' into their workflow, querying the assistant for galaxies similar to some galaxy of interest or guiding classification suggestions from the assistant.
Volunteers may discuss galaxies flagged by anomaly-detecting models, supporting serendipitous discovery by design.
At Galaxy Zoo, volunteers have always contributed far more than just answering the decision tree; we hope that algorithmic tools will complement and leverage their uniquely human skills.

\section{Conclusion}
\label{sec:conclusion}

We have presented detailed morphology measurements for 8.7M galaxies. Measurements include whether galaxies have bars (strong/weak/none), spiral arms (and their count and winding), and signs of recent mergers. We include measurements for the vast majority (97\%) of galaxies brighter than $r < 19$ in the DESI Legacy Surveys footprint (19,000 sq deg). We hope that these measurements will be useful for investigating how morphology and physical processes combine to create the diversity of galaxies we observe.

Our measurements are made by deep learning models trained on Galaxy Zoo volunteer responses from GZ DECaLS campaigns (GZD-1/2, GZD-5) as well as on newly-collected volunteer labels which we also share (GZD-8). 
Each campaign asked volunteers slightly different questions and received differently-distributed answers. To handle this, we introduce a novel loss function for learning from multiple campaigns simultaneously. This allows us to train on all responses and thereby create more capable models than we could from GZD-8 alone.
Our final models are trained on 401k galaxies labelled with 10M volunteer classifications.
They typically predict the fraction of volunteers selecting a given answer to within 5-10\% of the true fraction. 

Our measurements (both volunteer and automated) are available via \href{https://zenodo.org/record/7786417#.ZCcrqPdw3u0}{Zenodo}. Visible morphology changes significantly with increasing redshift and so we suggest that researchers control for this observational bias or focus on low-redshift subsets. 

Our models are available at \href{www.github.com/mwalmsley/zoobot}{GitHub} along with examples and documentation for adapting them to new datasets. We anticipate that these models will continue to evolve and we welcome collaborators.

Measuring the morphology of 8.7M galaxies with volunteers alone would have taken approximately 200 years. 
We achieved this scale by augmenting our volunteers with capable deep learning models.
This system will allow Galaxy Zoo to rapidly measure detailed morphology in all upcoming large-area surveys.




\bibliographystyle{mnras}
\bibliography{references}


\section*{Acknowledgements}
\label{sec:acknowledgements}

The data in this paper are the result of the efforts of the Galaxy Zoo volunteers, without whom none of this work would be possible. Their efforts are individually acknowledged at \href{http://authors.galaxyzoo.org}{http://authors.galaxyzoo.org}. We would also like to thank our volunteer translators; Mei-Yin Chou, Antonia Fernández Figueroa, Rodrigo Freitas, Na'ama Hallakoun, Lauren Huang, Alvaro Menduina, Beatriz Mingo, Verónica Motta, João Retrê, and Erik Rosenberg.

We would like to thank Dustin Lang for creating the \href{wwww.legacysurvey.org}{legacysurvey.org} cutout service and for contributing image processing code. We also thank Sugata Kaviraj and Matthew Hopkins for helpful discussions.

MW acknowledges funding from the Science and Technology Facilities Council (STFC) Grant Code ST/R505006/1.
We also acknowledge support from STFC under grant ST/N003179/1. LF and KBM acknowledge partial support from the US National Science Foundation award IIS 2006894. 
BDS acknowledges support through a UK Research and Innovation Future Leaders Fellowship [grant number MR/T044136/1]. ILG acknowledges support from an STFC PhD studentship [grant number ST/T506205/1]. DOR acknowledges support from an STFC PhD studentship [grant number ST/T506205/1]. ILG and DOR also acknowledge support from the Faculty of Science and Technology at Lancaster University.

This publication uses data generated via the Zooniverse.org platform, development of which is funded by generous support, including a Global Impact Award from Google, and by a grant from the Alfred P. Sloan Foundation.

This research made use of the open-source Python scientific computing ecosystem, including SciPy \citep{Jones2001}, Matplotlib \citep{Hunter2007}, scikit-learn \citep{Pedregosa2012}, scikit-image \citep{VanderWalt2014} and Pandas \citep{McKinney2010}. This research made use of Astropy, a community-developed core Python package for Astronomy \citep{TheAstropyCollaboration2018}. This research made use of PyTorch \citep{Paszke2019}, Pyro \citep{bingham2018pyro}, TensorFlow \citep{Abadi2015}, and TensorFlow Probability \citep{Dillon2017}.

The Legacy Surveys consist of three individual and complementary projects: the Dark Energy Camera Legacy Survey (DECaLS; NSF's OIR Lab Proposal ID \# 2014B-0404; PIs: David Schlegel and Arjun Dey), the Beijing-Arizona Sky Survey (BASS; NSF's OIR Lab Proposal ID \# 2015A-0801; PIs: Zhou Xu and Xiaohui Fan), and the Mayall z-band Legacy Survey (MzLS; NSF's OIR Lab Proposal ID \# 2016A-0453; PI: Arjun Dey). DECaLS, BASS and MzLS together include data obtained, respectively, at the Blanco telescope, Cerro Tololo Inter-American Observatory, The NSF's National Optical-Infrared Astronomy Research Laboratory (NSF's OIR Lab); the Bok telescope, Steward Observatory, University of Arizona; and the Mayall telescope, Kitt Peak National Observatory, NSF's OIR Lab. The Legacy Surveys project is honored to be permitted to conduct astronomical research on Iolkam Du'ag (Kitt Peak), a mountain with particular significance to the Tohono O'odham Nation.
The NSF's OIR Lab is operated by the Association of Universities for Research in Astronomy (AURA) under a cooperative agreement with the National Science Foundation.
This project used data obtained with the Dark Energy Camera (DECam), which was constructed by the Dark Energy Survey (DES) collaboration. Funding for the DES Projects has been provided by the U.S. Department of Energy, the U.S. National Science Foundation, the Ministry of Science and Education of Spain, the Science and Technology Facilities Council of the United Kingdom, the Higher Education Funding Council for England, the National Center for Supercomputing Applications at the University of Illinois at Urbana-Champaign, the Kavli Institute of Cosmological Physics at the University of Chicago, Center for Cosmology and Astro-Particle Physics at the Ohio State University, the Mitchell Institute for Fundamental Physics and Astronomy at Texas A\&M University, Financiadora de Estudos e Projetos, Fundacao Carlos Chagas Filho de Amparo, Financiadora de Estudos e Projetos, Fundacao Carlos Chagas Filho de Amparo a Pesquisa do Estado do Rio de Janeiro, Conselho Nacional de Desenvolvimento Cientifico e Tecnologico and the Ministerio da Ciencia, Tecnologia e Inovacao, the Deutsche Forschungsgemeinschaft and the Collaborating Institutions in the Dark Energy Survey. The Collaborating Institutions are Argonne National Laboratory, the University of California at Santa Cruz, the University of Cambridge, Centro de Investigaciones Energeticas, Medioambientales y Tecnologicas-Madrid, the University of Chicago, University College London, the DES-Brazil Consortium, the University of Edinburgh, the Eidgenossische Technische Hochschule (ETH) Zurich, Fermi National Accelerator Laboratory, the University of Illinois at Urbana-Champaign, the Institut de Ciencies de l'Espai (IEEC/CSIC), the Institut de Fisica d'Altes Energies, Lawrence Berkeley National Laboratory, the Ludwig-Maximilians Universitat Munchen and the associated Excellence Cluster Universe, the University of Michigan, the National Optical Astronomy Observatory, the University of Nottingham, the Ohio State University, the University of Pennsylvania, the University of Portsmouth, SLAC National Accelerator Laboratory, Stanford University, the University of Sussex, and Texas A\&M University.
BASS is a key project of the Telescope Access Program (TAP), which has been funded by the National Astronomical Observatories of China, the Chinese Academy of Sciences (the Strategic Priority Research Program `The Emergence of Cosmological Structures' Grant \# XDB09000000), and the Special Fund for Astronomy from the Ministry of Finance. The BASS is also supported by the External Cooperation Program of Chinese Academy of Sciences (Grant \# 114A11KYSB20160057), and Chinese National Natural Science Foundation (Grant \# 11433005).
The Legacy Survey team makes use of data products from the Near-Earth Object Wide-field Infrared Survey Explorer (NEOWISE), which is a project of the Jet Propulsion Laboratory/California Institute of Technology. NEOWISE is funded by the National Aeronautics and Space Administration.
The Legacy Surveys imaging of the DESI footprint is supported by the Director, Office of Science, Office of High Energy Physics of the U.S. Department of Energy under Contract No. DE-AC02-05CH1123, by the National Energy Research Scientific Computing Center, a DOE Office of Science User Facility under the same contract; and by the U.S. National Science Foundation, Division of Astronomical Sciences under Contract No. AST-0950945 to NOAO.

For the purpose of open access, the author has applied a Creative Commons Attribution (CC BY) licence to any Author Accepted Manuscript version arising.

\appendix

\section{Data Availability}
\label{appendix:data_availability}

The data underlying this article are available via Zenodo at \href{https://doi.org/10.5281/zenodo.7786416}{https://doi.org/10.5281/zenodo.7786416}. Any future data updates will be released using Zenodo versioning - please check you are viewing the latest version. 
\\
\\
The automated morphology catalogue is also available at CDS via anonymous ftp to cdsarc.u-strasbg.fr (130.79.128.5) or via \href{https://cdsarc.unistra.fr/viz-bin/cat/J/MNRAS}{https://cdsarc.unistra.fr/viz-bin/cat/J/MNRAS}. CDS offers \href{http://cdsportal.u-strasbg.fr/}{tools} including a TAP/ADQL server and the \href{https://vizier.cds.unistra.fr/viz-bin/VizieR}{VizieR} catalog viewer.\footnote{For this arXiv version, the catalogue may not be immediately available via VizieR}
\\
\\
The catalogue is also available at NOIRLab's Astro Data Lab via \href{https://datalab.noirlab.edu/galaxy_zoo_desi}{https://datalab.noirlab.edu/galaxy\_zoo\_desi/}. Astro Data Lab offers \href{https://datalab.noirlab.edu/tools.php}{tools} including an ADQL web interface and remotely hosted Jupyter notebooks.\footnote{For this arXiv version, the catalogue may not be immediately available via NOIRLab}
\\
\\
The code and trained models underlying this article are available at \href{https://github.com/mwalmsley/zoobot}{https://github.com/mwalmsley/zoobot}.

\section{Field-of-View Selection}
\label{sec:field-of-view}

GZ DECaLS images had a field-of-view calculated using the 50\% and 90\% Petrosian radii reported by the NASA-Sloan Atlas photometry pipeline applied to SDSS images (see W+22 Eqn. 1). SDSS images and NSA photometry are not available outside the SDSS footprint and so cannot be used to set the field-of-view directly. However, 614k galaxies (96\%) of galaxies in the NSA cross-match to extended sources in the DESI-LS catalogue (10 arcsec matching radius) and hence have both NSA and \texttt{tractor} photometry. We use these galaxies to estimate the NSA Petrosian radii from \texttt{tractor} photometry. 

\begin{figure}
    \centering
    \includegraphics[width=\columnwidth]{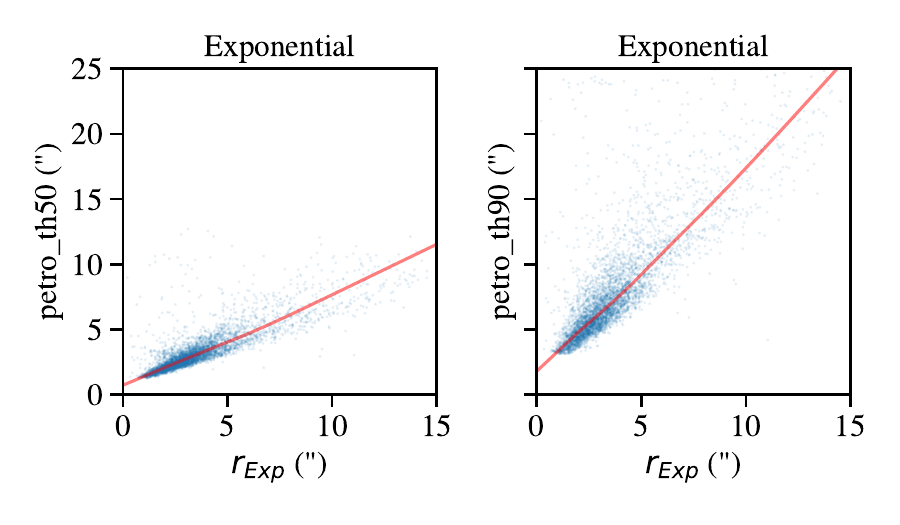}
    \includegraphics[width=\columnwidth]{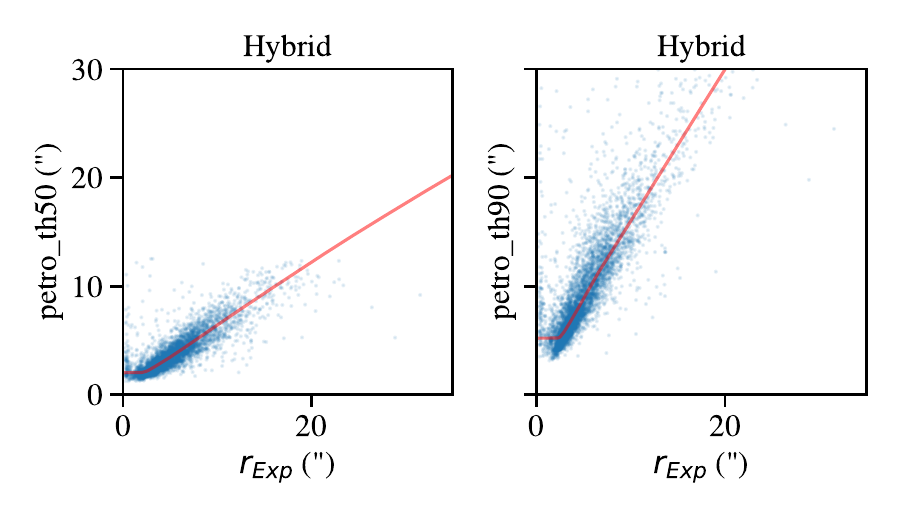}
    \caption{Estimating SDSS/NSA radii measurements from DESI-L/\texttt{tractor} measurements, for \texttt{tractor} sources reported as having exponential-dominated or hybrid light profiles. $r_{\rm Exp}$ is the half-light radius of the r-band exponential profile reported by \texttt{tractor}, while \texttt{petro\_th50} and \texttt{petro\_th90} are the 50\% and 90\% Petrosian radii reported by the NSA. Blue points are galaxies with both \texttt{tractor} and NSA measurements, and red curves are the resulting nonparametric LOWESS fit (local sampling fraction of 0.3).}
    \label{fig:exp_other}
\end{figure}

\begin{figure}
    \centering
    \includegraphics[width=\columnwidth]{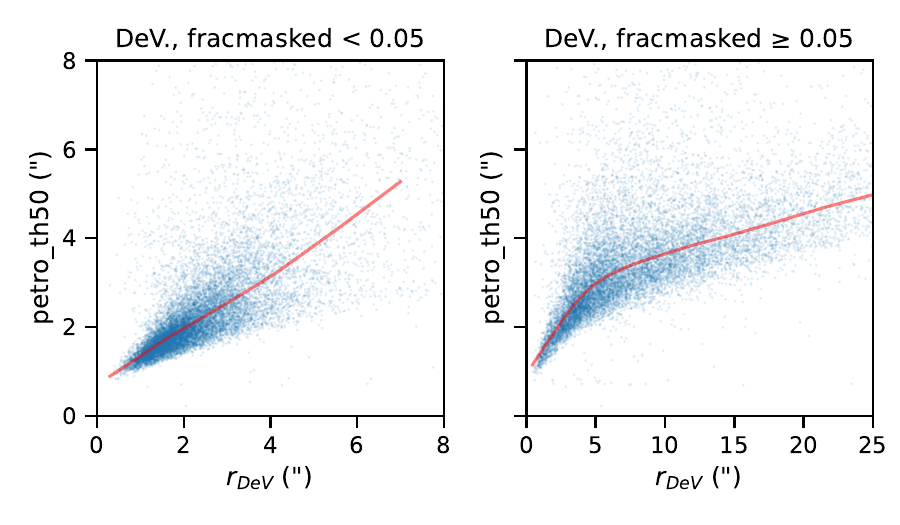}
    \includegraphics[width=\columnwidth]{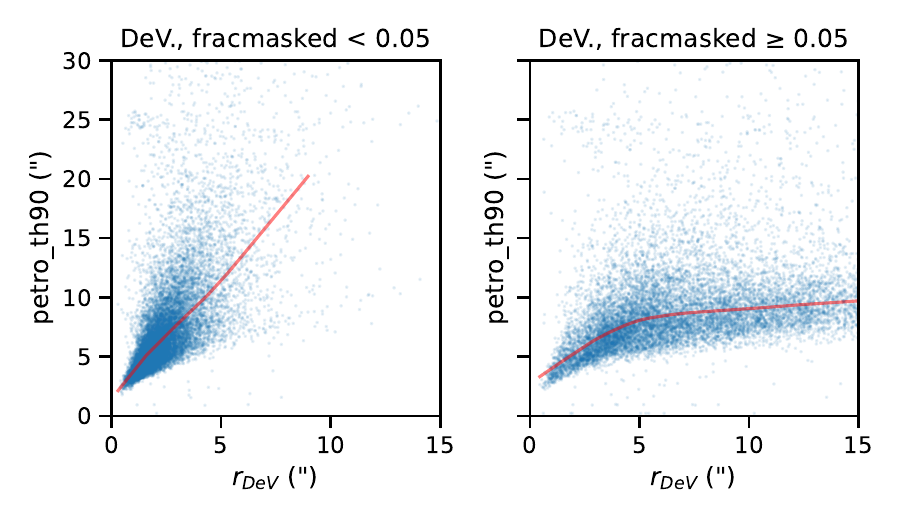}
    \caption{As with Fig. \ref{fig:exp_other}, but for \texttt{tractor} sources reported as having De Vaucouleurs (DeV) dominated light profiles. $r_{\rm DeV}$ is the half-light radius of the r-band De Vaucouleurs profile reported by \texttt{tractor}. Empirical relationships are fit separately for galaxies with over 5\% \texttt{tractor}-masked pixels (\texttt{fracmasked}), as these follow a qualitatively different relationship. }
    \label{fig:dev_fracmasked}
\end{figure}

Recall (Sec \ref{sec:source_identification}) that the DESI-LS sourcefinding algorithm \texttt{tractor} models sources as a weighted mixture of exponential and De Vaucouleurs light profiles.
Most sources are well-described by a single profile (which we define as at least 95\% of the flux accounted for); 55\% are DeV-dominated, 32\% are Exp-dominated, and the remaining 12\% are hybrid.
For each population (DeV-dominated, Exp-dominated, and hybrid), we fit the relationship between the \texttt{tractor}-measured shape and the NSA-measured Petrosian radii (50\% and 90\%). We then use these empirical relationships to estimate the Petrosian radii which would have been measured for galaxies with only \texttt{tractor} photometry, and hence to estimate the field-of-view which GZ DECaLS would have used. Fig. \ref{fig:exp_other}-\ref{fig:dev_fracmasked} shows these empirical relationships.
Note that these relationships can only be reliably estimated where sufficient data exists; we discard 30,551 galaxies (0.03\%) with extreme \texttt{tractor}-reported sizes.

\begin{figure}
    \centering
    \includegraphics[width=\columnwidth]{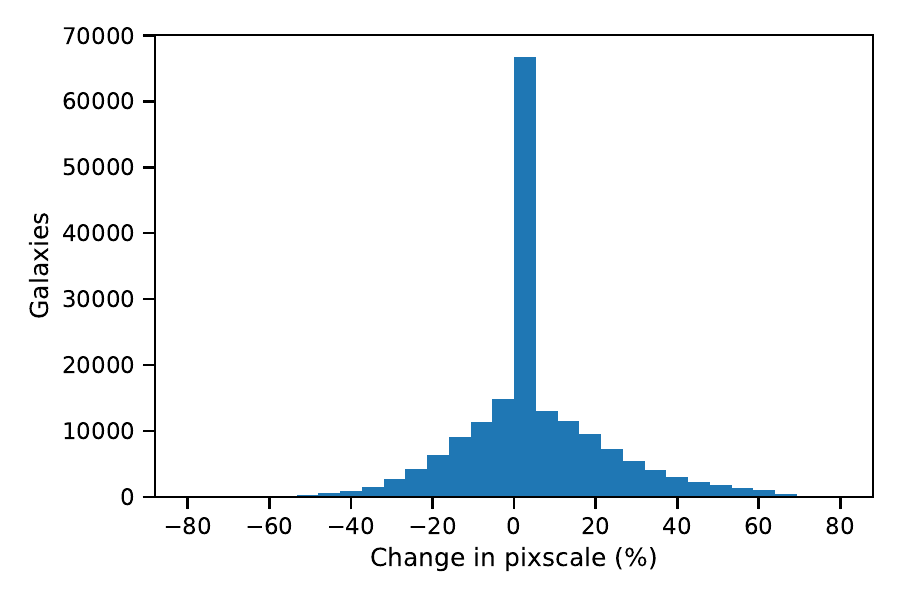}
    \caption{Change in resized pixel scale (pixscale), proportional to field-of-view in the final resized RGB images . The vast majority of galaxies have a \texttt{tractor}-estimated pixel scale within 20\% of the pixel scale they would have had if we had access to NSA Petrosian radii. They would therefore have similar fields-of-view to GZ DECaLS galaxies, ensuring our DECaLS-derived labels remain applicable.}
    \label{fig:change_in_pixscale}
\end{figure}

Fig. \ref{fig:change_in_pixscale} shows the percentage change in field-of-view calculated with our empirical estimates of NSA radii from \texttt{tractor} photometry vs. directly from NSA radii. The clear majority of galaxies are within 20\% of the field-of-view that would have been used in GZ DECaLS. Our new DESI-LS images will therefore have similar fields-of-view to the previously-labelled GZ DECaLS images, ensuring our DECaLS-derived labels remain applicable.

\section{Model and Training Details}
\label{sec:training_details}

Subsections \ref{sec:loss} and \ref{sec:additional_labels} describe our core advances, namely, a new multi-campaign loss and new volunteer responses as labels.
This Appendix documents additional details which are not novel advances but are important for reproducibility and of potential interest to the specialized reader.

\subsubsection{Training Catalogues}

Our models are trained on galaxies drawn from three volunteer-labelled catalogues, corresponding to three Galaxy Zoo campaigns - GZD-1/2, GZD-5, and GZD-8. This subsection includes practical notes on the construction of these catalogues.

We require that galaxies only appear in a single catalogue, to avoid the possibility of duplicated galaxies (and hence train/test contamination) when training on multiple catalogues.
Galaxies that were classified in multiple campaigns are assigned to the catalogue of the campaign in which they received the most responses.
Every galaxy is listed exactly once and, when listed, it has responses from every campaign it appeared in.
31,747 galaxies appear\footnote{As recorded by their NASA-Sloan Atlas `iauname' identifier.} in both GZD-1/2 and GZD-5. Of those, 24,633 have more total votes in GZD-1/2 and are placed in the GZD-1/2 training catalogue, while the remaining 7,114 galaxies have more total votes in GZD-5 and are placed in the GZD-5 training catalogue. When selecting galaxies to upload for GZD-8, we removed any galaxies listed in the \texttt{tractor} catalogue (Sec. \ref{sec:data}) within $5\arcsec$ of a galaxy already uploaded in GZD-1/2 or GZD-5, and hence there are no GZD-8 galaxies classified in earlier campaigns.
Table \ref{tab:counts} reports the galaxies and vote counts for each training catalogue.

\begin{table}
        
        \begin{tabular}{lllrr}
        \toprule
         Catalogue & Galaxies & GZD-1/2 Votes & GZD-5 Votes & GZD-8 Votes\\
        \midrule
        GZD-1/2 & 82k &  3.26M & 175k & 0 \\
        GZD-5 & 223k & 253k & 3.83M & 0 \\
        GZD-8 & 96k & 0 & 0 & 3.06M \\
        \bottomrule
        \end{tabular}
        \caption{Counts of galaxies and volunteer votes present in our GZD-1/2, GZD-5, and GZD-8 training catalogues. Galaxies classified in more than one campaign are placed in the training catalogue for the campaign in which they received the most votes. All votes from all campaigns are kept, and hence some GZD-1/2 galaxies have some votes collected during GZD-5 and vice versa.}
     \label{tab:counts}
\end{table}

\subsubsection{Model Details}

Our base model is EfficientNetB0. EfficientNetB0 is the smallest (by parameter count) version of the EfficientNet architecture family introduced by \cite{Tan2019a}. W+22 Sec. 5.1 provides an astronomer-orientated introduction to the enhancements made vs. prior architectures. \cite{Fielding2021} benchmarked the performance of our code and found EfficientNet competitive with other architectures. As in W+22, we replace EfficientNetB0's default head with a single dense layer of one unit per answer, each with a sigmoid activation to ensure Dirichlet-appropriate outputs between 1--101. Unlike W+22, we separately predict answers to every GZ DESI campaign (Sec. \ref{sec:loss}) and hence we have 98 output units.  We adjust the EfficientNetB0 weights using the Adam optimiser \citep{Kingma2015} with the hyperparameters below (Sec. \ref{sec:hyperparameter_search}). We experimented with reducing the learning rate on loss plateau and found no convincing evidence that this improved performance. 

We train our models using 2 40GB NVIDIA A100 GPUs available via IRIS\footnote{https://www.iris.ac.uk/}. We train using \texttt{PyTorch}'s `distributed data parallel' configuration i.e. each model draws subbatches from a different fixed data split and shares weight updates. Training time for each model depends on the dataset chosen. Training on GZD-5 (223k galaxies) takes approx. 6 hours while training on all campaigns (GZD-1/2/5/8, 401k galaxies) takes approx. 10 hours.

\subsubsection{Hyperparameter Search}
\label{sec:hyperparameter_search}

The hyperparameters (and fundamental design) of our base model, EfficientNetB0 \citep{Tan2019a}, were chosen to optimise a balance of prediction FLOPs and test accuracy on ImageNet \citep{Russakovsky2015}. The ideal hyperparameters may vary by task and dataset, and so it is plausible that better hyperparameters exist for our specific goal of predicting Galaxy Zoo volunteer votes. In this Appendix, we search for those better hyperparameters, and find that the published hyperparameters are indeed close to optimal for our specific task.

It is important to avoid the possibility that our reported performance improvements from training on all campaigns (Sec. \ref{sec:results}) are caused by the hyperparameters of the model simply being highly tuned to learning from all campaigns.
To avoid this, we only tune our hyperparameters using the GZD-5 campaign data.\\

Our hyperparameter search procedure is as follows.
We first select our hyperparameters to optimise, dividing them into architectural or augmentation hyperparameters. For our architectural parameters, we select the image size (as interpolated and input to the network), the batch size, the learning rate, the drop-connect rate, the (head) dropout rate, and the $\beta_0$ momentum parameter of the Adam optimiser. For our augmentation parameters, we select the upper and lower bounds of the relative crop size and the upper and lower bounds of the cropped aspect ratio. We then execute a random search for each set of hyperparameters i.e. we train many models with randomised architectural or augmentation hyperparameters. We train 152 models with randomised architectural parameters and 55 models with randomised augmentation parameters, reflecting our larger architectural search space. Random searches are robust and effective when the important hyperparameters are not previously known \cite{Bergstra2012}. We assume that the best choice of architectural parameters is independent of the best choice of augmentation parameters, and so they can be searched separately.

When training each model, we divide GZD-5 into random train/validation/test splits of size 70\%/10\%/20\%. We train the model until the validation loss does not decrease for 10 epochs (early stopping) and then record the test loss. Optimising the test loss is appropriate because the validation loss will be biased low due to early stopping, and because we will make no further changes to the model design before retraining on additional campaigns and reporting our performance.

\begin{figure}
    \centering
    \includegraphics{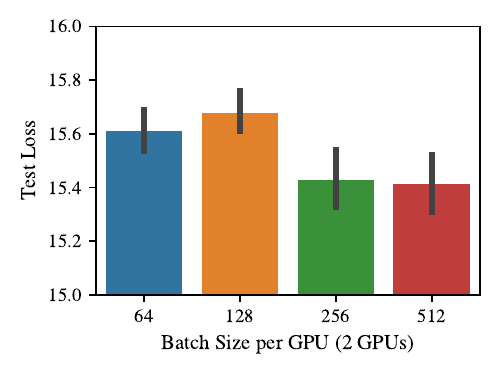}
    \includegraphics{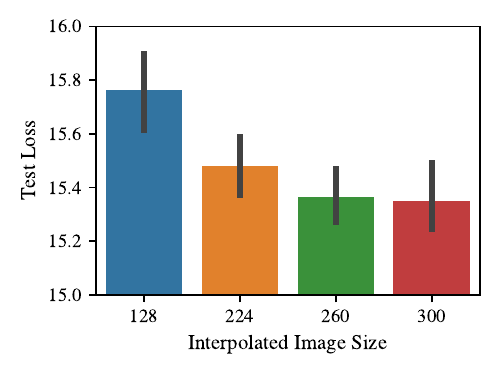}
    \includegraphics{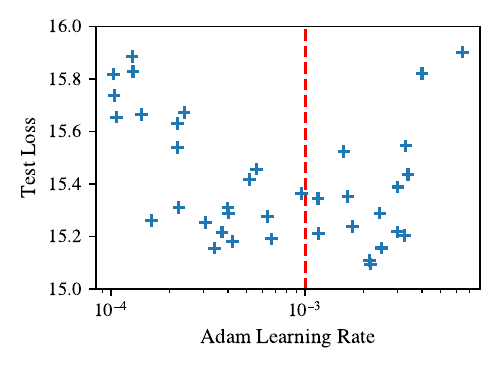}
    \caption{Architectural hyperparameter search results for batch size (upper), image size (middle), and learning rate (lower). Larger batch sizes and larger images are likely helpful. The ideal learning rate is likely close to the conventional $10^{-3}$ value. 
    \\
    \\
    Test loss given a hyperparameter is calculated after filtering for models where the other hyperparameters are close to optimal i.e. after fixing batch size to 128 or 256, and/or image size to 224 or 260. Errorbars show the 95\% confidence interval on the mean test loss.}
    \label{fig:hparam_vs_loss}
\end{figure}

Fig. \ref{fig:hparam_vs_loss} shows the effect of key architectural hyperparameters.
We find that larger batch sizes and larger image sizes likely improve performance.
We find that the ideal learning rate is likely close to the conventional default of $10^{-3}$ (which we ultimately opt to use).
We find no significant evidence that the drop-connect rate, dropout rate or $\beta_0$ momentum affect performance on our task, and set these to their conventional defaults (0.2, 0.5 and 0.9 respectively). 

We ultimately select a batch size of 256 (512 across both GPUs) and an image size of 224. 
We felt that the additional memory footprint of training on larger images was not a sensible trade-off.
Our goal is to create models which other researchers can easily finetune.
The image size during finetuning must match our (pre)training image size and so (pre)training on larger images would significantly increase the memory footprint required to use our models (e.g. +60\% for 300x300 vs. 224x224 images)
We prefer to create models which have slightly less-than-optimal performance but are practical for other researchers to use. 
 
For the augmentation hyperparameters, we find no significant evidence that either relative crop size and cropped aspect ratio affect test performance, and so we arbitrarily set these hyperparameters to visually sensible values (0.7\textendash 0.8 relative crop size bounds and 0.9\textendash 1.1 cropped aspect ratio bounds).

\section{GZD-8 Confusion Matrices}
\label{sec:confusion_appendix}

This Appendix shows the confusion matrices for each question for a random model trained on all campaigns (GZD-1/2/5/8) making predictions on a random 20\% test subset of GZD-8.

Our models predict posteriors for the expected distribution of possible responses.
Here, for intuition only, these posteriors are converted to discrete classifications by rounding the observed vote fraction (label) and mean of the expected vote count posterior (prediction) to the nearest integer. The matrices then show the counts of rounded predictions (x axis) against rounded labels (y axis). To avoid the loss of information from rounding, we encourage researchers not to treat Galaxy Zoo classifications as discrete, and instead to use the full vote fractions or posteriors where possible. 

We define a correct classification as one where the answer with the highest predicted vote fraction matches the answer with the highest actual volunteer vote fraction. We also apply the minimum total votes of 34 and relevance criteria described in Sec. \ref{sec:breakdown}).

\begin{figure*}
    
    \includegraphics[width=.5\textwidth]{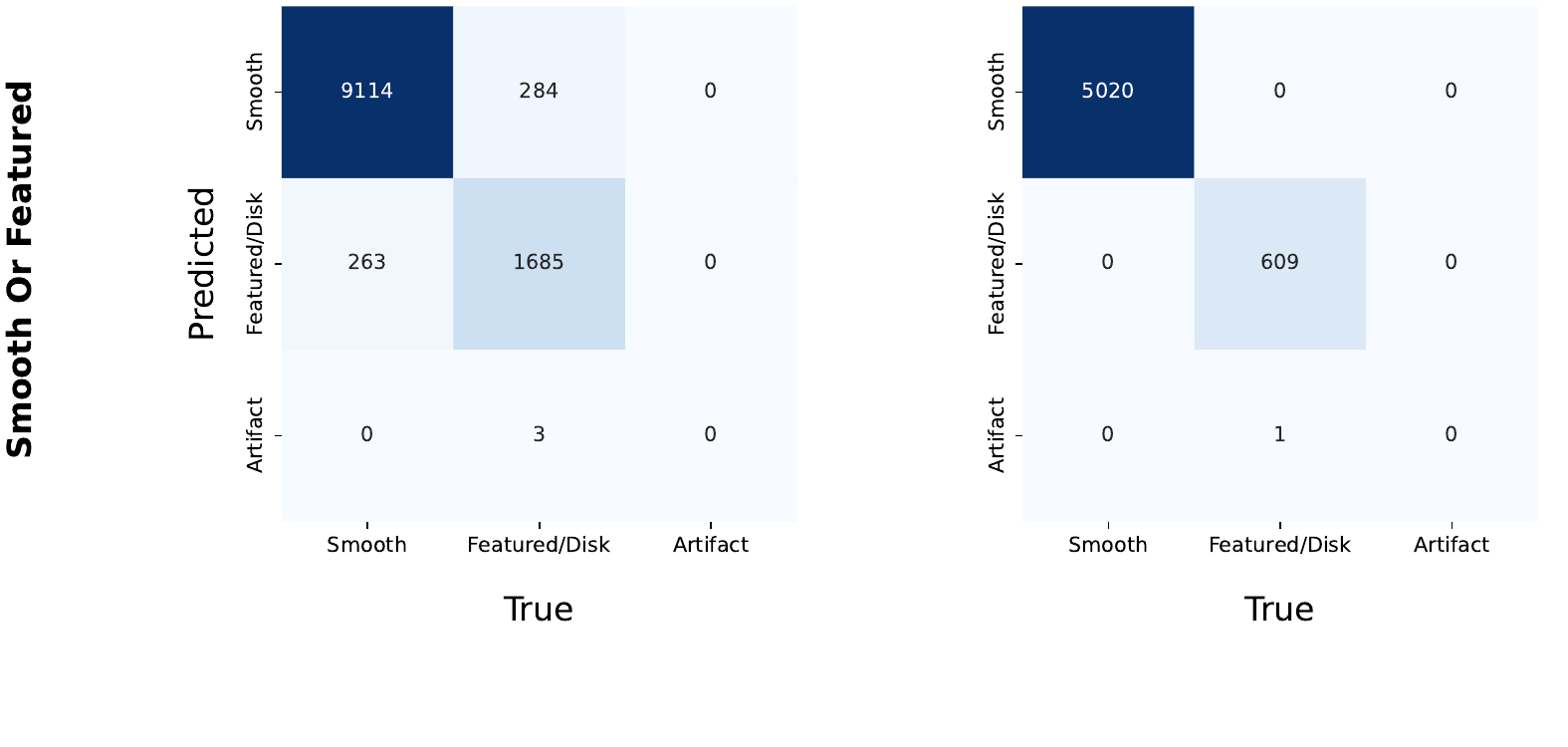}
        
    \includegraphics[width=.5\textwidth]{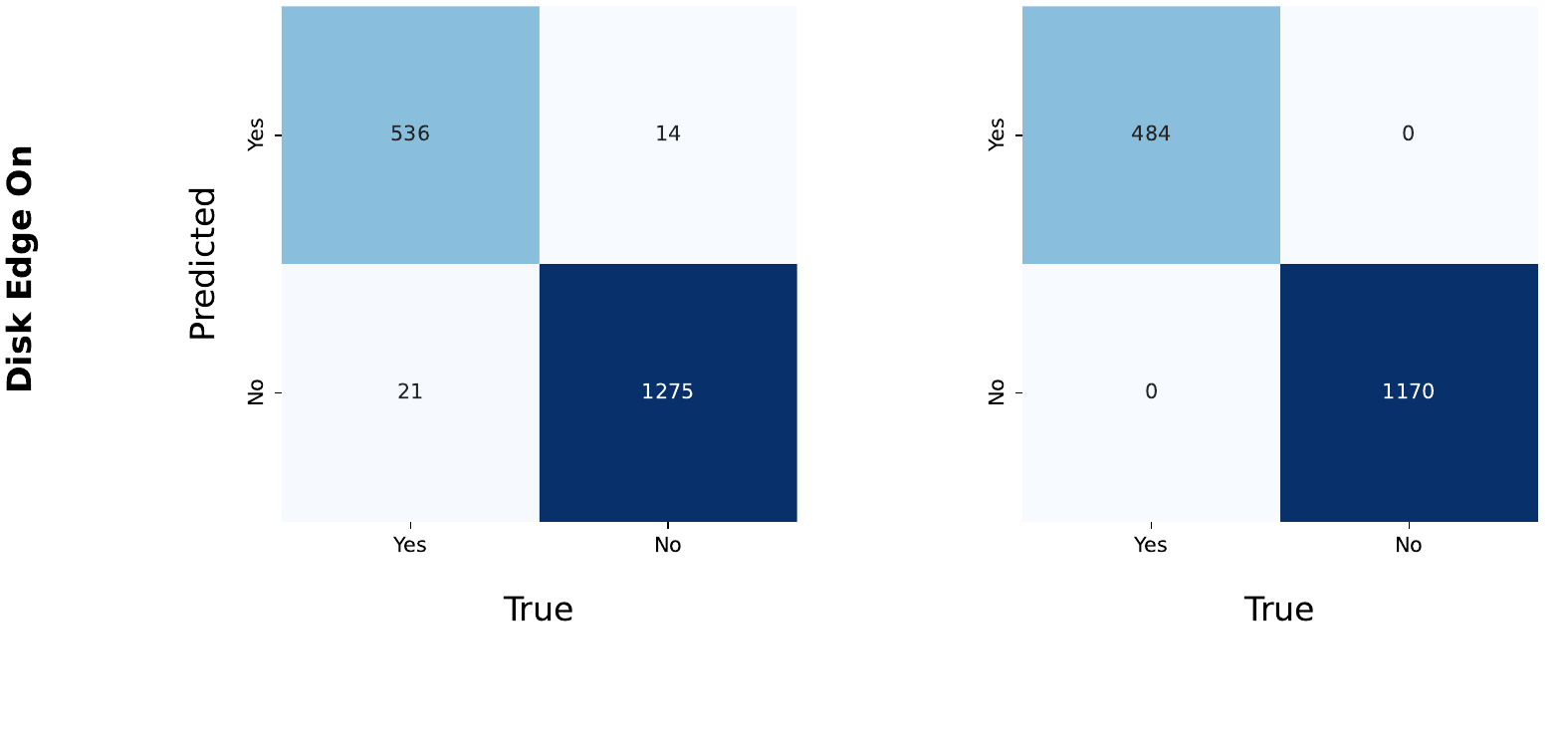}
    
    \includegraphics[width=.5\textwidth]{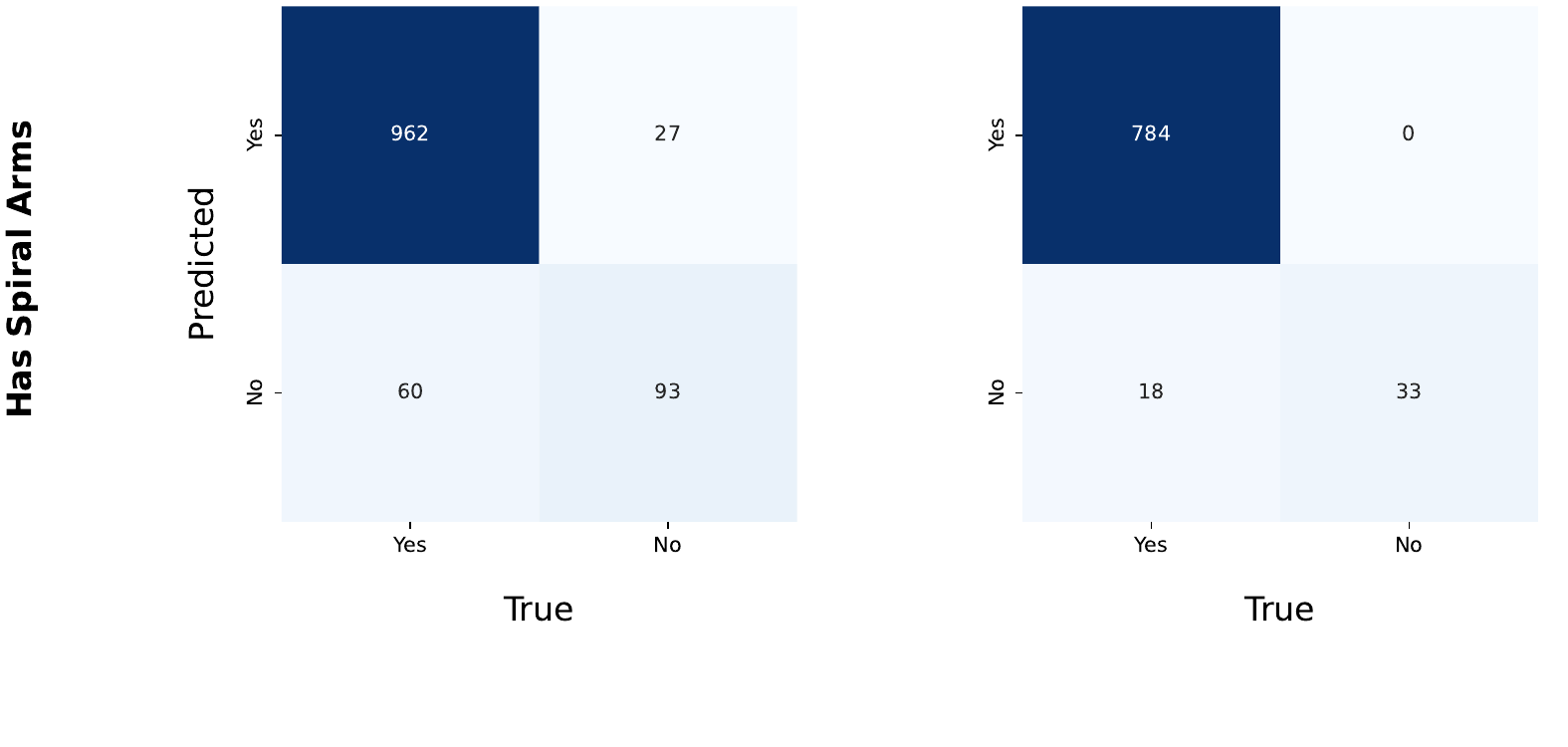}
    
    \includegraphics[width=.5\textwidth]{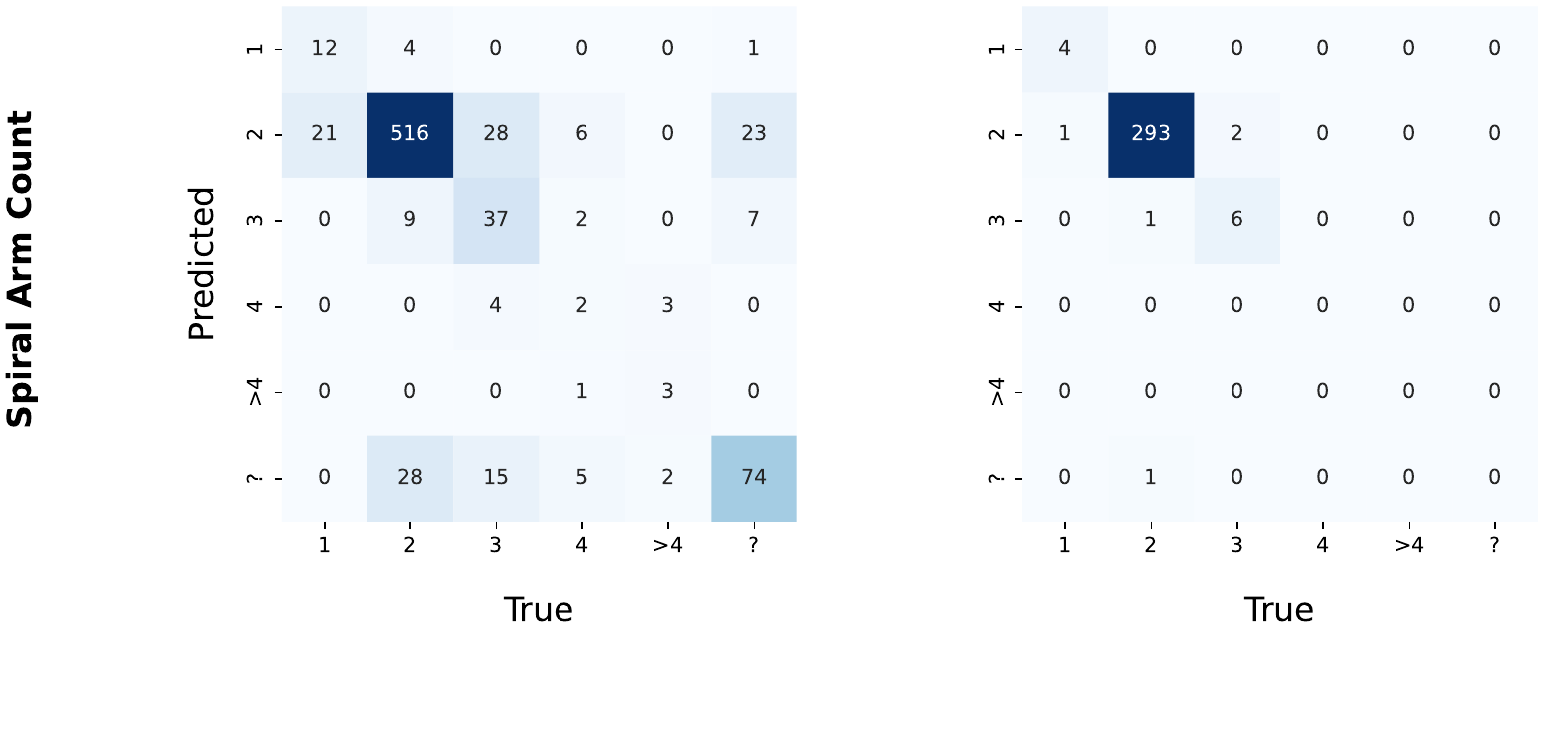}
    
    \includegraphics[width=.5\textwidth]{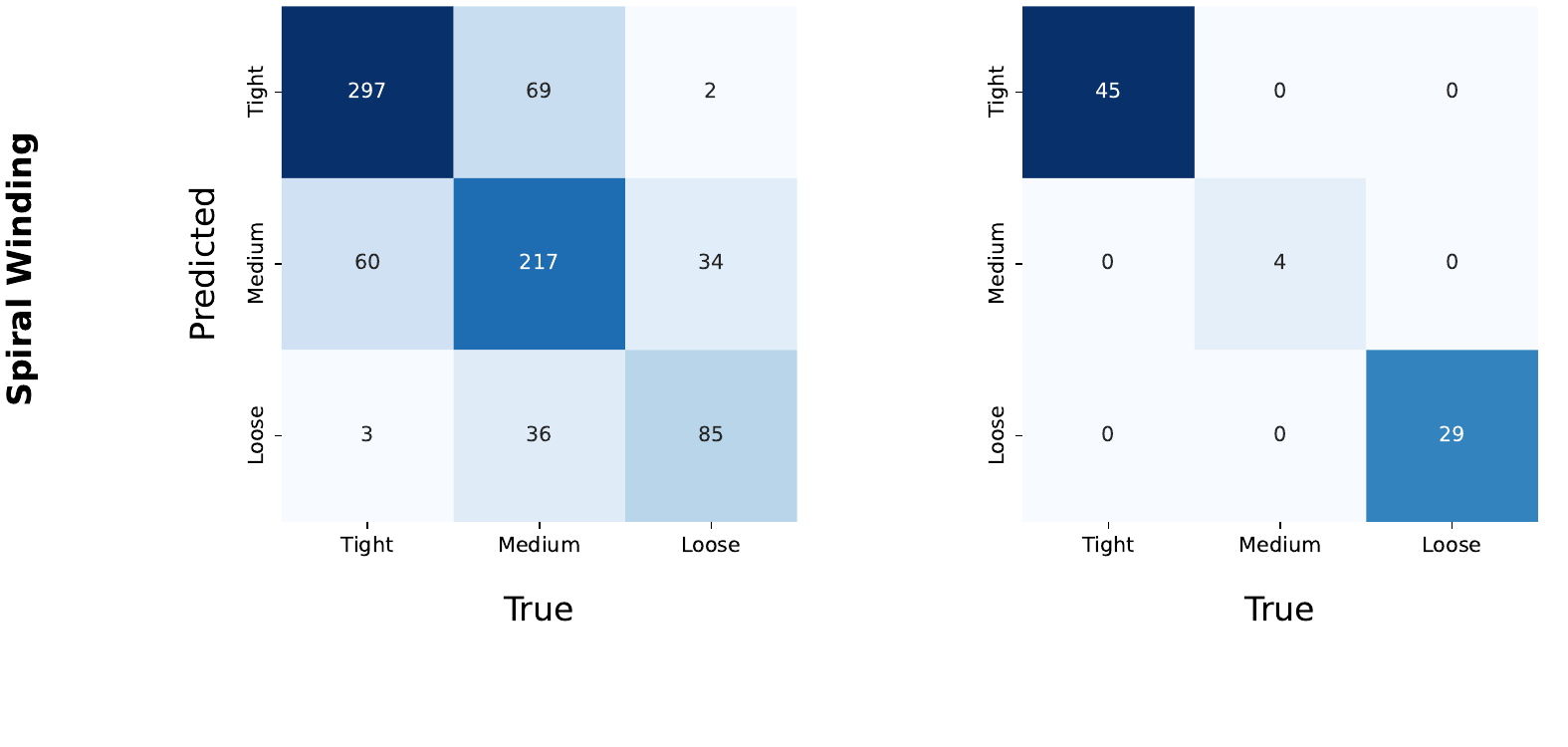}
    
    \hspace{52pt}\textit{All Galaxies} \hspace{52pt} \textit{High Volunteer Confidence}
    
    \caption{
        Confusion matrices for each question, made on the 11,349 galaxies in the (random) GZD-8 test set with at least 34 votes. Classifications are considered correct if the answer with the highest predicted vote fraction matches the answer with the highest actual volunteer vote fraction.
        For each question's confusion matrix, we only show galaxies where a majority of volunteers were asked that question (see main text).
        The right-hand matrices are additionally filtered to only show galaxies where the volunteers were confident. This is defined as the actual volunteer vote fraction being greater than 0.8 i.e. where at least 80\% of volunteers agreed on an answer.
        }
    \label{fig:confusion_matrices}
  \end{figure*}

\begin{figure*}

    \includegraphics[width=.5\textwidth]{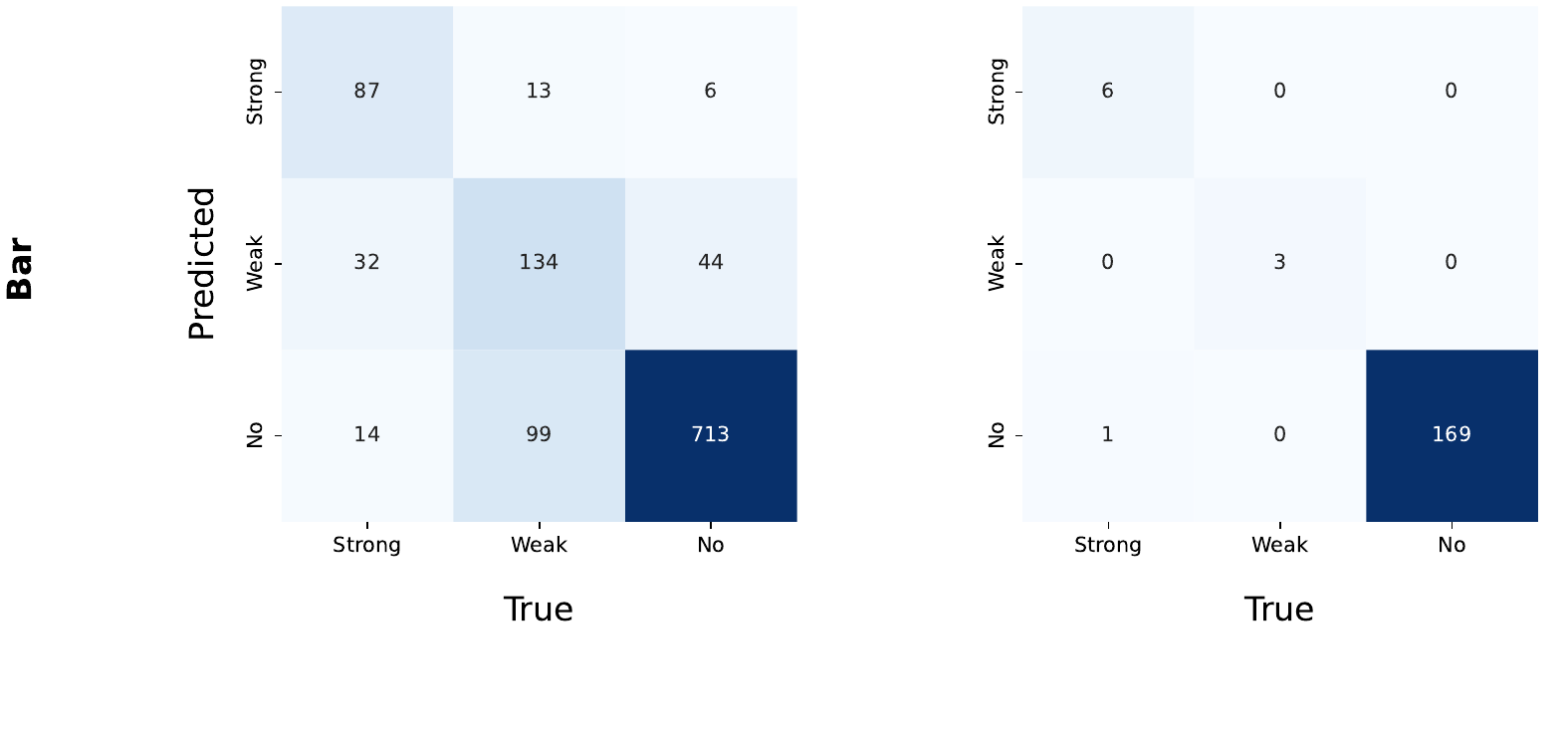}
        
    \includegraphics[width=.5\textwidth]{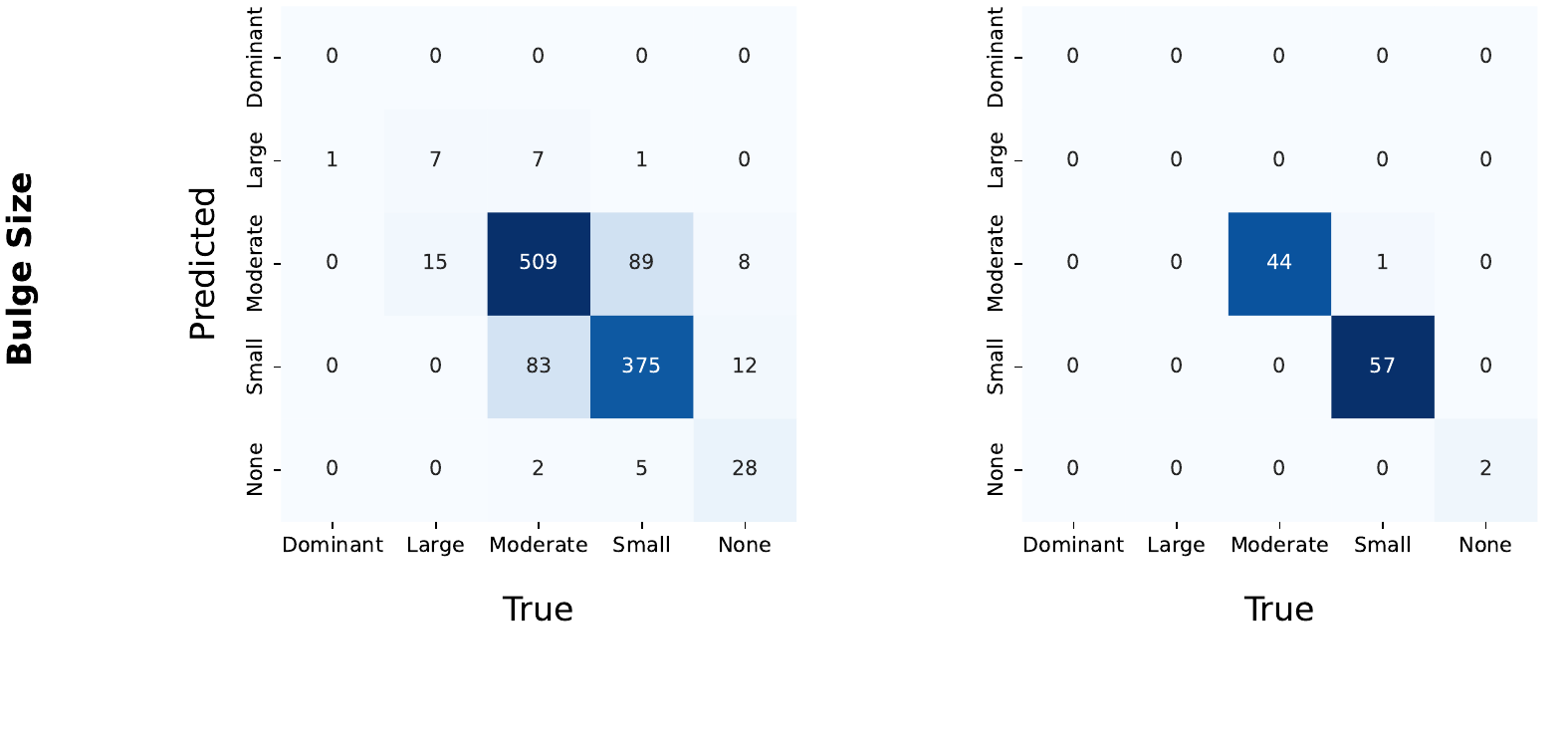}
    
    \includegraphics[width=.5\textwidth]{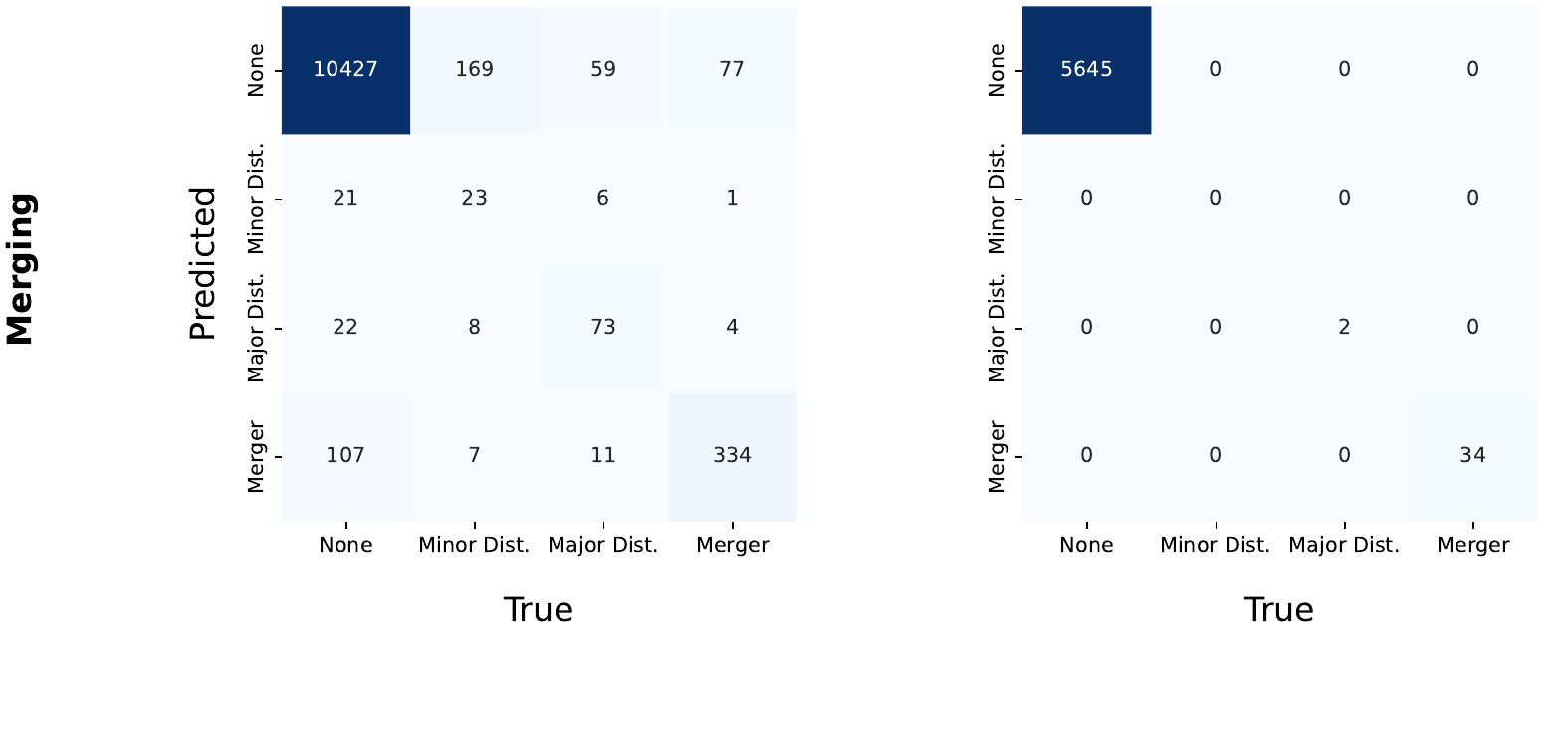}
    
    \includegraphics[width=.5\textwidth]{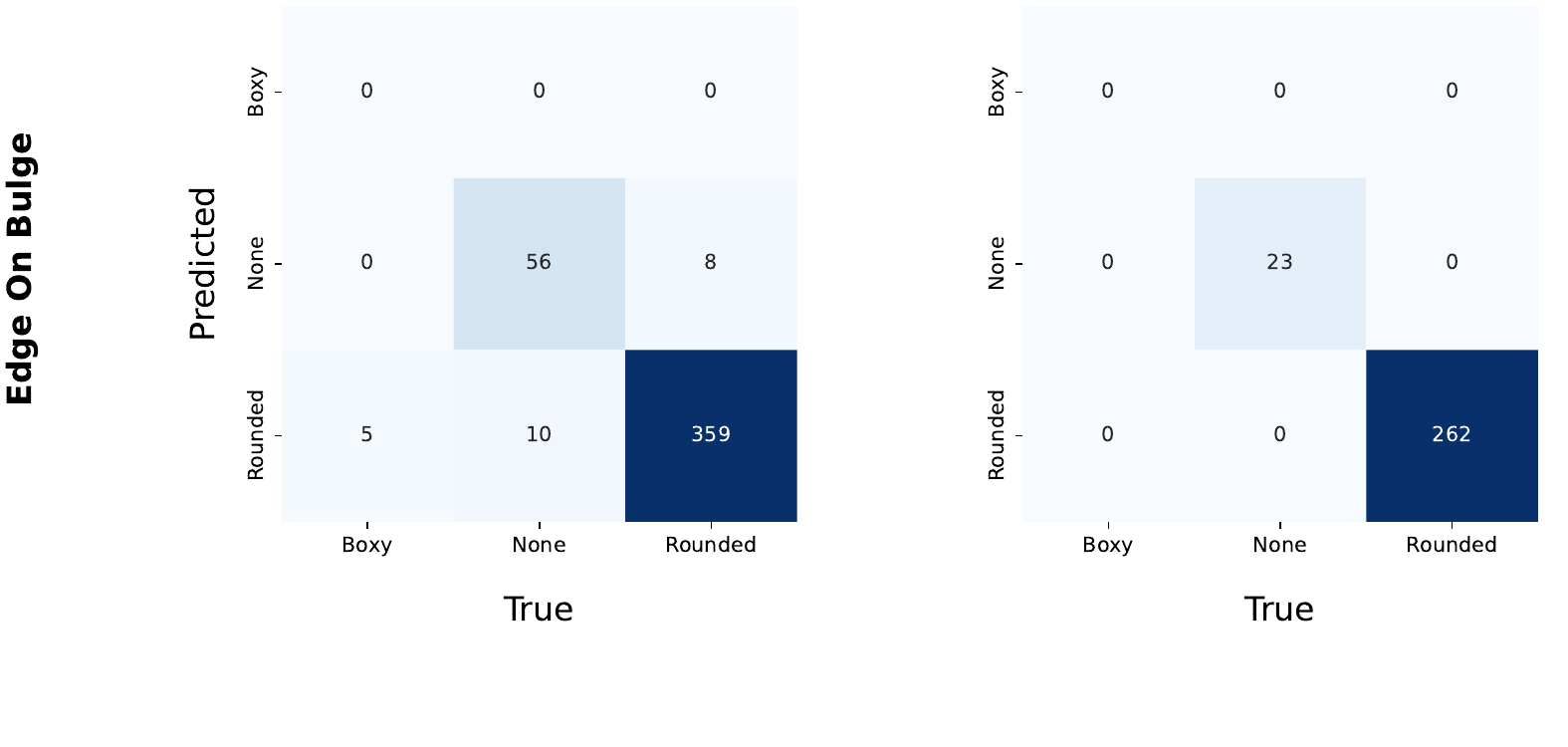}
    
    \includegraphics[width=.5\textwidth]{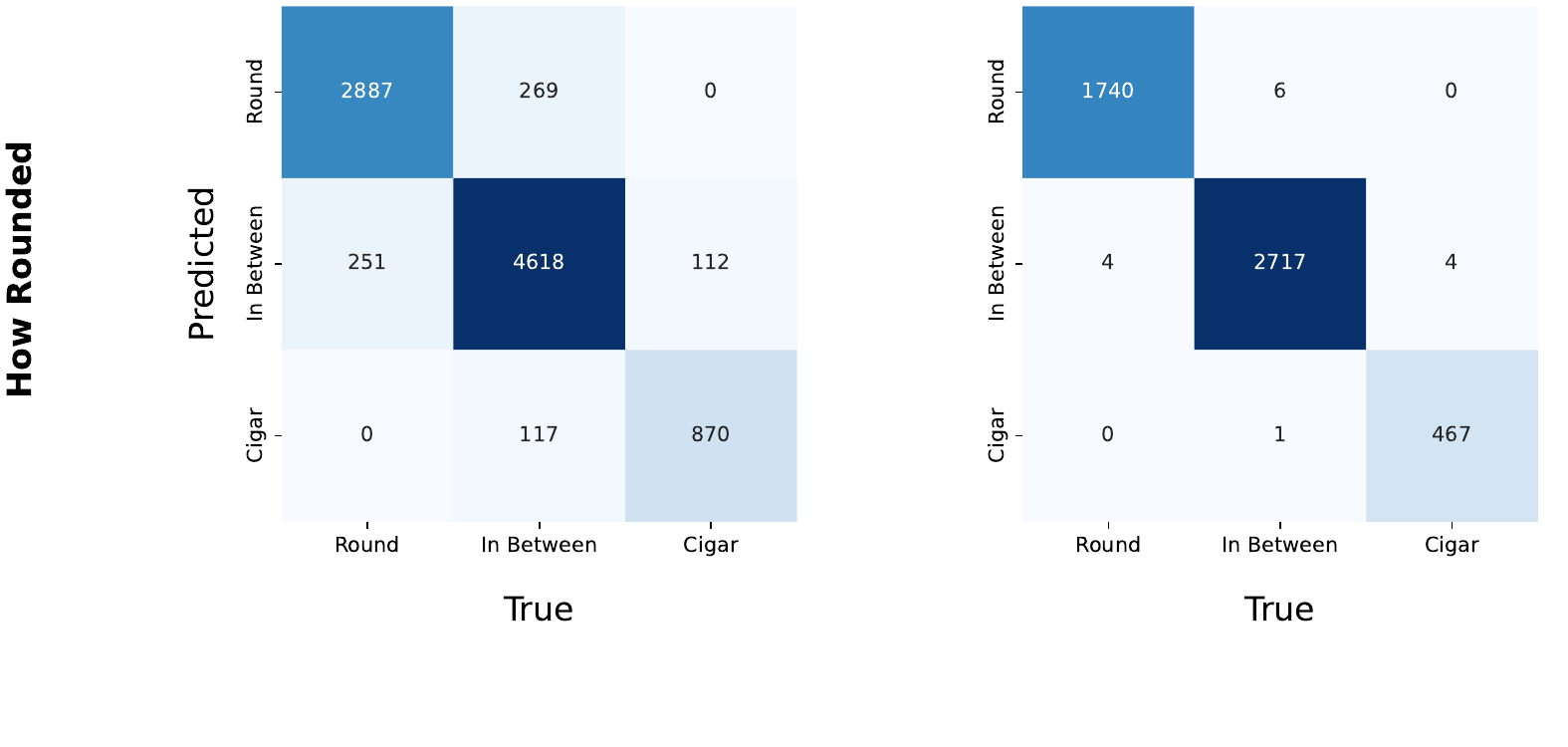}

    \hspace{52pt}\textit{All Galaxies} \hspace{52pt} \textit{High Volunteer Confidence}

    \caption{Continuing Fig. \ref{fig:confusion_matrices} above.}
    \label{fig:confusion_matrices_2}
\end{figure*}

Fig. \ref{fig:confusion_matrices}-\ref{fig:confusion_matrices_2} shows confusion matrices for all galaxies passing the total votes and relevance criteria above (left) and for only those galaxies where volunteers were confident (right), defined as the actual volunteer vote fraction being greater than 0.8.

\section{Cross-Matching to External Data}
\label{sec:crossmatching}

Galaxy morphology is one of many measurable galaxy properties. 
We have cross-matched our GZ DESI morphology catalogue to several external catalogues of other galaxy properties.
We hope that this will help reveal how morphology affects, and is affected by, those properties.

We include data from the NASA-Sloan Atlas (NSA, \citealt{Aguado2019}), the OSSY Type 1 AGN catalogue \citep{Oh2015}, the Arecibo Legacy Fast ALFA Survey (ALFALFA, \citealt{Haynes2018}), the MPA-JHU SDSS DR7 derived properties catalogue \citep{Abazajian2009}, and the DESI photometric redshift catalogue by \citep{Zhou2021}. Please credit the original authors of these external catalogues when using their data.

We cross-match the quoted (optical) coordinates of sources in each of the above catalogues with the \texttt{tractor} source catalogue coordinates underlying our morphology catalogue. We match sources within 10 arcseconds. For the rare case where multiple external sources match a \texttt{tractor} source, the closest external source is selected and any remaining sources are dropped.

We combine the redshifts from these external catalogues for Fig. \ref{fig:frac_vs_redshift} and when selecting the low redshift subsets in our data release.
We select redshifts in the following priority order: SDSS spectroscopic redshifts from the NSA, then from OSSY, and then the spec\_z and photo\_z columns from \cite{Zhou2021}.

\bsp	
\label{lastpage}
\end{document}